\documentclass[preprints, article,accept,moreauthors,pdftex,10pt,a4paper]{mdpi}
\usepackage{aas_macros}

\firstpage{1} 
\pubvolume{xx}
\issuenum{1}
\articlenumber{1}
\pubyear{2018}
\copyrightyear{2018}
\history{Received: date; Accepted: date; Published: date}

\Title{Opacity Effects on Pulsations of Main-Sequence A-Type Stars}

\Author{Joyce A. Guzik $^{1}$, Christopher J. Fontes $^{2}$ and Chris Fryer $^{3}$}

\AuthorNames{Joyce A. Guzik, Christopher J. Fontes and Chris Fryer}

\address{%
$^{1}$ \quad Los Alamos National Laboratory, Los Alamos, NM; joy@lanl.gov\\
$^{2}$ \quad Los Alamos National Laboratory, Los Alamos, NM; cjf@lanl.gov\\
$^{3}$ \quad Los Alamos National Laboratory, Los Alamos, NM; fryer@lanl.gov}

\abstract{Opacity enhancements for stellar interior conditions have been explored to explain observed pulsation frequencies and to extend the pulsation instability region for B-type main-sequence variable stars \citep[see, e.g.,][]{2015A&A...580L...9W, 2017arXiv170101258W, 2017MNRAS.466.2284D, 2016MNRAS.455L..67M, 2014A&A...565A..76C,2009AIPC.1170..388Z}.  For these stars, the pulsations are driven in the region of the opacity bump of Fe-group elements at $\sim$200,000 K in the stellar envelope.  Here we explore effects of opacity enhancements for the somewhat cooler main-sequence A-type stars, in which $p$-mode pulsations are driven instead in the second helium ionization region at $\sim$50,000 K. We compare models using the new LANL OPLIB \citep{2016ApJ...817..116C} vs. LLNL OPAL \citep{1996ApJ...464..943I} opacities for the AGSS09 \cite{2009ARA&A..47..481A} solar mixture.  For models of 2 solar masses and effective temperature 7600 K, opacity enhancements have only a mild effect on pulsations, shifting mode frequencies and/or slightly changing kinetic-energy growth rates.  Increased opacity near the bump at 200,000 K can induce convection that may alter composition gradients created by diffusive settling and radiative levitation.  Opacity increases around the hydrogen and 1st He ionization region (13,000 K) can cause additional higher-frequency $p$ modes to be excited, raising the possibility that improved treatment of these layers may result in prediction of new modes that could be tested by observations.  New or wider convective zones and higher convective velocities produced by opacity increases could also affect angular momentum transport during evolution.  More work needs to be done to quantify the effects of opacity on the boundaries of the pulsation instability regions for A-type stars.}


\begin{document}


\section{Introduction}

Opacities, element abundances, and convection affect pulsation instabilities in nearly all types of stars. Studies of pulsating B-type stars, namely the $\beta$ Cep pressure-mode pulsators, SPB (Slowly Pulsating B) gravity-mode pulsators, and their hybrids that pulsate in both types of modes, have motivated opacity investigations. For these stars, pulsations are driven by the $\kappa$ (opacity valving) mechanism \citep{1980tsp..book.....C} in the stellar envelope around 200,000 K, where the opacity is increased by bound-bound transitions in Fe-group elements (the so-called Z-bump).  SPB/$\beta$ Cep stars show fewer modes than expected, and hybrids that are not predicted \cite[see, e.g.,][re.~$\nu$ Eri and 12 Lac]{2008MNRAS.385.2061D}.  Increasing the Fe/Ni opacities by a factor of 1.75, as inferred from Sandia pulsed-power experimental data \citep{2015Natur.517...56B}, widens the hybrid instability region enough to include 12 Lac \citep{2016MNRAS.455L..67M}.  Opacity modifications improve the match to pulsation observations for $\nu$ Eri \citep{2017MNRAS.466.2284D}.  The SPB and $\beta$ Cep pulsation-instability regions are somewhat wider \citep{2015A&A...580L...9W} using the new Los Alamos OPLIB \citep{2016ApJ...817..116C} opacities.

The ``solar abundance problem'' also motivates opacity increases.  Standard solar models using the latest abundance determinations  and taking into account uncertainties in input physics show disagreement between the sound speed inferred from solar $p$ modes and that of the standard model \citep{2017ApJ...835..202V}.  The sound-speed discrepancy becomes slightly smaller using the new LANL OPLIB opacities instead of the LLNL OPAL opacities \citep{2016ApJ...817..116C, 2016IAUFM..29B.532G}.  Inversions for solar entropy \citep{2017A&A...607A..58B} and the Ledoux discriminant \citep{2017MNRAS.472L..70B} give additional constraints on the solar interior, highlighting further the effects of opacity differences between OPAL and OPLIB.

For A-type stars, namely the $\delta$ Sct variables, $p$-mode pulsations are driven by the $\kappa$ effect produced by the 2nd ionization of helium around 50,000 K.  Balona et al. \citep{2015MNRAS.452.3073B} carried out an extensive parameter study to search for pulsation instability of low-frequency modes that are found in many $\delta$ Sct variables observed by the {\it Kepler} spacecraft.  They varied mass, helium abundance, metallicity, rotation rate, and considered OPAL vs. OP \citep{2005MNRAS.362L...1S} opacities.  They also consider opacity enhancements around the Z-bump (log T = 5.35, or T = 224,000 K), and at the so-called ``Kurucz bump'' at log T = 5.06 (T = 114,800 K), first noted in opacities developed for model atmospheres \citep{2003IAUS..210P.A20C} and discussed by Cugier \cite{2014A&A...565A..76C} for B-type star pulsations.  Balona et al. \cite{2015MNRAS.452.3073B} find that an opacity enhancement of at least a factor of two at the Kurucz bump is needed to excite low-frequency modes; however, they do not advocate such an opacity enhancement as the explanation for the low frequencies observed in $\delta$ Sct stars.

Daszy{\'n}ska-Daszkiewicz et al. \cite{2017EPJWC.16003013D} explored the effects of opacity enhancements to explain the 16 to 115 $\mu$Hz variability of the rapidly rotating Maia variables, with spectral types B8 to A2, between the $\delta$ Sct and SPB variables.  They explored opacity enhancements for 2.5-3.2 M$_\odot$ rotating models, finding that a factor of 1.5 enhancement around log T = 5.1 (T = 126,000 K), near the Kurucz bump, produces pulsation instability in the observed frequency range.

Here we further explore the consequences of opacities and opacity enhancements for A-type stars.  We compare 2 M$_\odot$ evolution and pulsation models using the new LANL OPLIB vs. the LLNL OPAL opacities.  We examine the effects of opacity multipliers in the region of bumps at 15,000, 50,000, and 2.6 million K, in addition to the Z-bump region around 200,000 K.  We use different stellar evolution and pulsation codes than those used in previous studies.  We show the effects of opacity enhancements on inducing and widening convection zones and increasing convective velocities, in the context of the mixing-length theory of convection \citep{1958ZA.....46..108B}.  We estimates the effects of turbulent broadening on line opacities that could contribute to opacity enhancements.
 
\section{Stellar Evolution and Pulsation Models}

We use as our testbed for opacity studies the evolution of 2 M$_\odot$ models with OPAL \citep{1996ApJ...464..943I} and OPLIB opacities \citep{2016ApJ...817..116C}, and the AGSS09 \cite{2009ARA&A..47..481A} solar abundance mixture.  See \cite{2016IAUFM..29B.532G} and references therein for details of the stellar evolution and nonadiabatic pulsation codes used for these studies.  We used mixing-length to pressure scale-height ratio 2.0 and did not include convective overshoot or rotation in our models. 

\subsection{LLNL OPAL vs. LANL OPLIB opacities}

\begin{figure}
\center
    \includegraphics[width=0.4\textwidth]{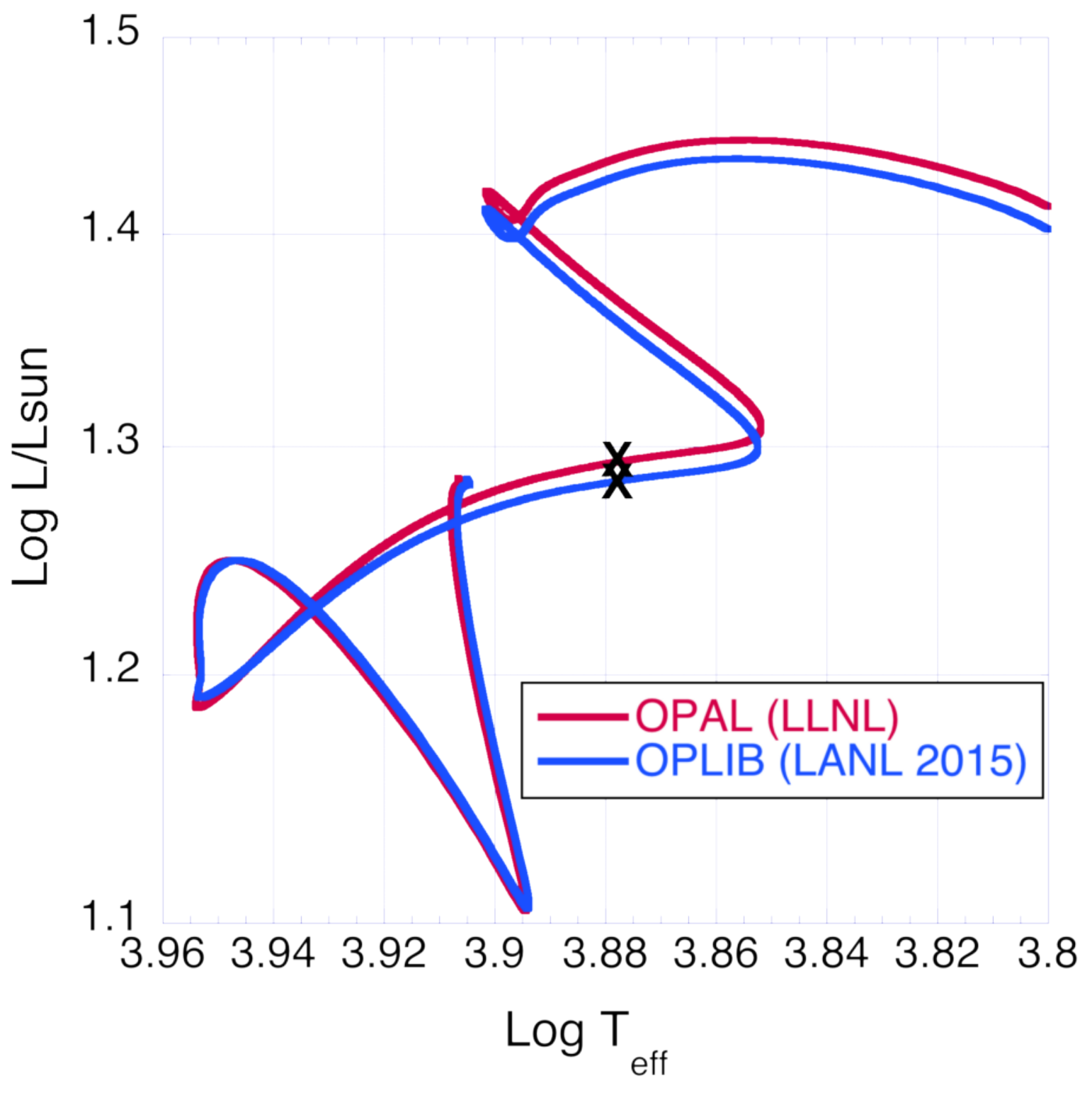}
     \includegraphics[width=0.4\textwidth]{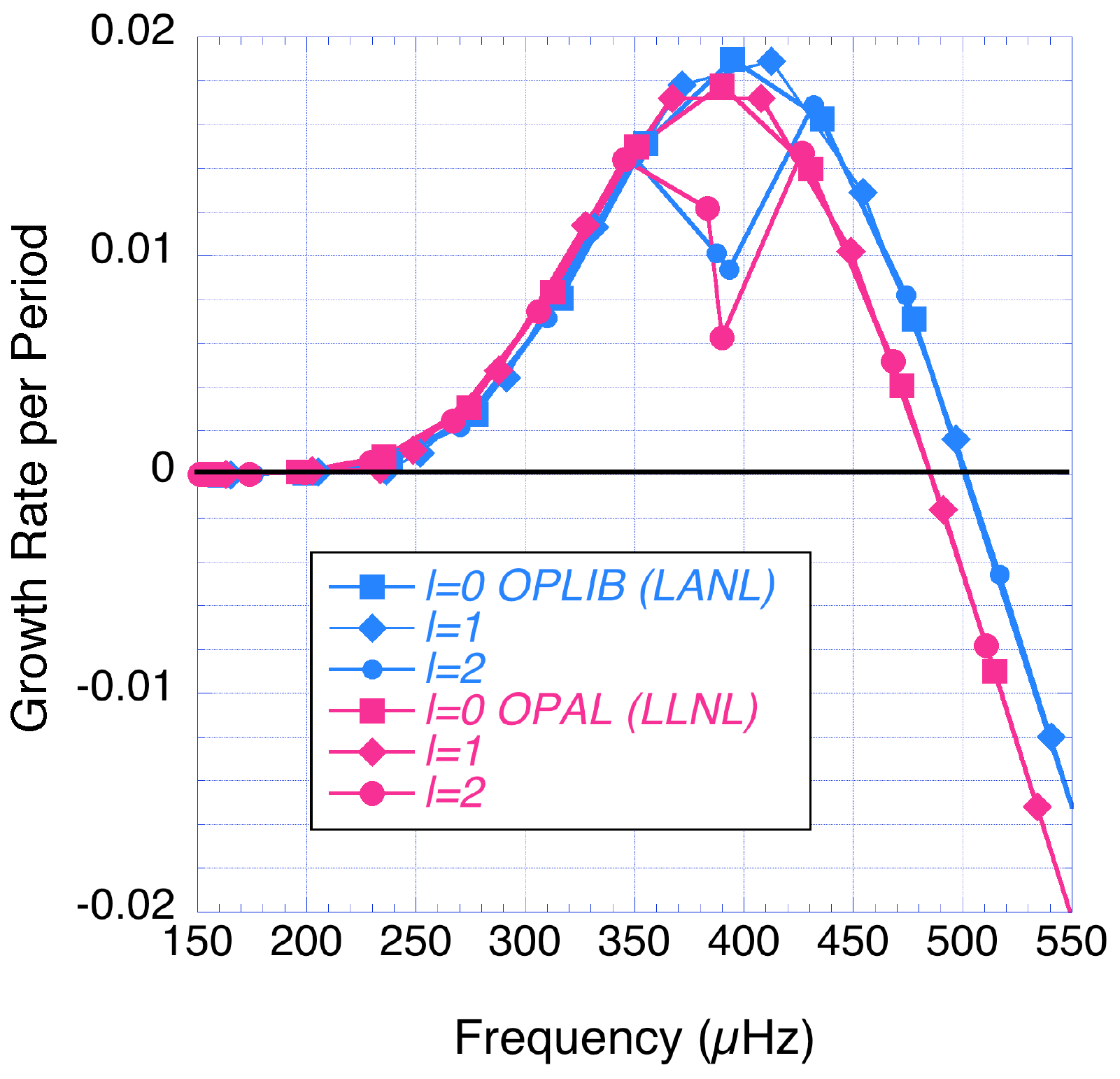}
    \caption{Evolution tracks in H-R diagram for 2 M$_\odot$ models with OPAL and OPLIB opacities (left); fractional kinetic-energy growth rate per period (dimensionless) vs.~frequency for $l$=0, 1, and 2 $p$ modes for T$_{\rm eff}$ = 7600 K models (right).  The $\bf{X}$s mark the model locations on the H-R diagram.}
    \label{fig:OPLIBvsOPAL}
\end{figure}

We first examine the difference in evolution track in the Hertzsprung-Russell (H-R) diagram (log luminosity vs. log effective temperature) and the calculated frequencies and growth rates of degree $l$=0-2 $p$ modes for T$_{\rm eff}$ = 7600 K models using OPAL vs.~OPLIB opacities (Fig. \ref{fig:OPLIBvsOPAL}).  The 2 M$_\odot$ model using OPAL opacities evolves at slightly higher luminosity for the same effective temperature. We found slightly higher growth rates at for the $l$=0-2 $p$ modes using the OPLIB opacities.

\subsection{$\times$2 Z-bump at 200,000 K}

\begin{figure}
\center
    \includegraphics[width=0.4\textwidth]{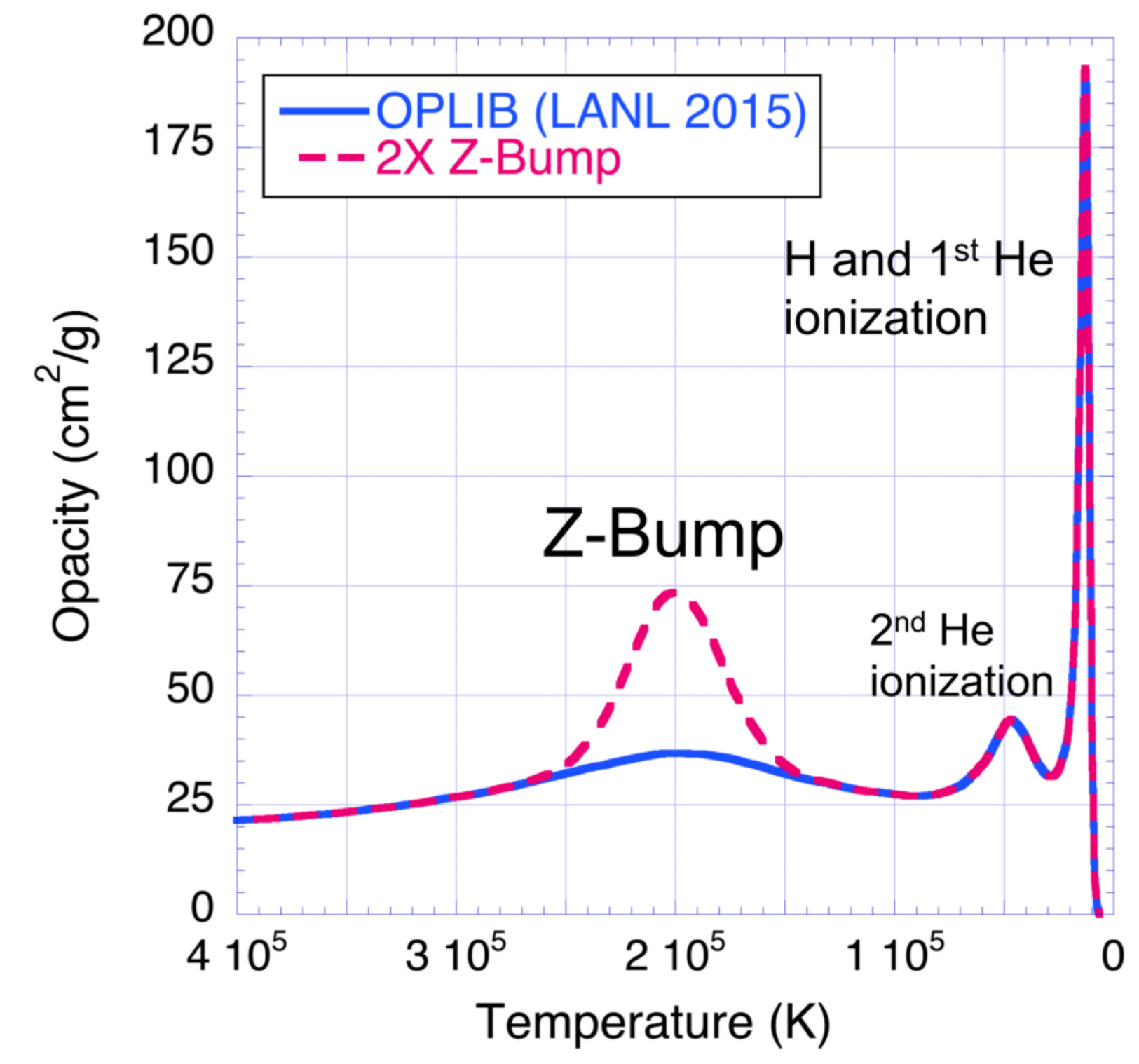}
     \includegraphics[width=0.4\textwidth]{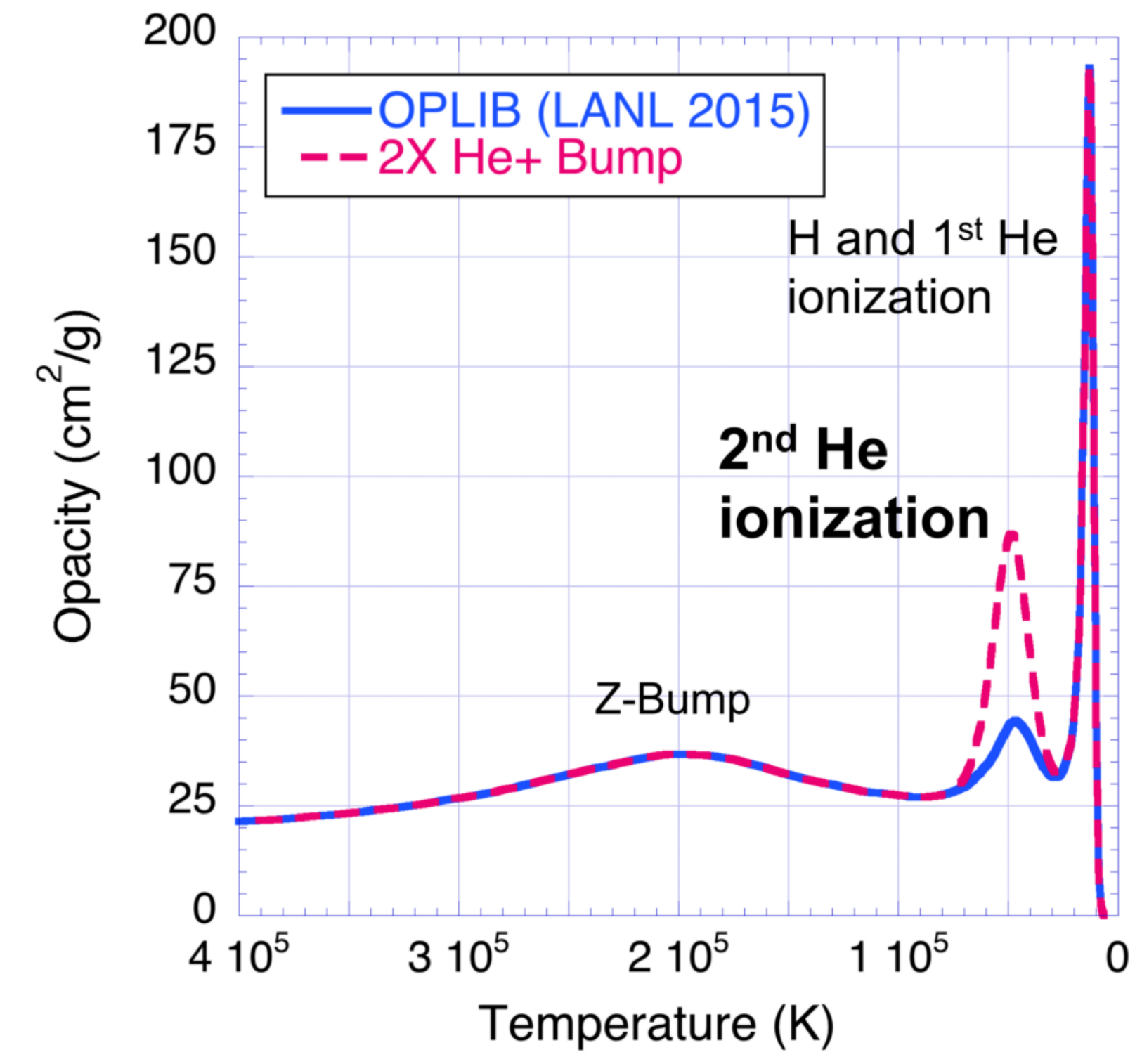}
    \caption{Opacity vs. stellar interior temperature for 2 M$_\odot$ models with T$_{\rm eff}$  = 7600 K, showing modification of the Z-bump opacity at 200,000 K (left) and of the 2nd He ionization region opacity at 50,000 K (right).}
    \label{fig:Opacity2xZbump}
\end{figure}

\begin{figure}
\center
    \includegraphics[width=0.4\textwidth]{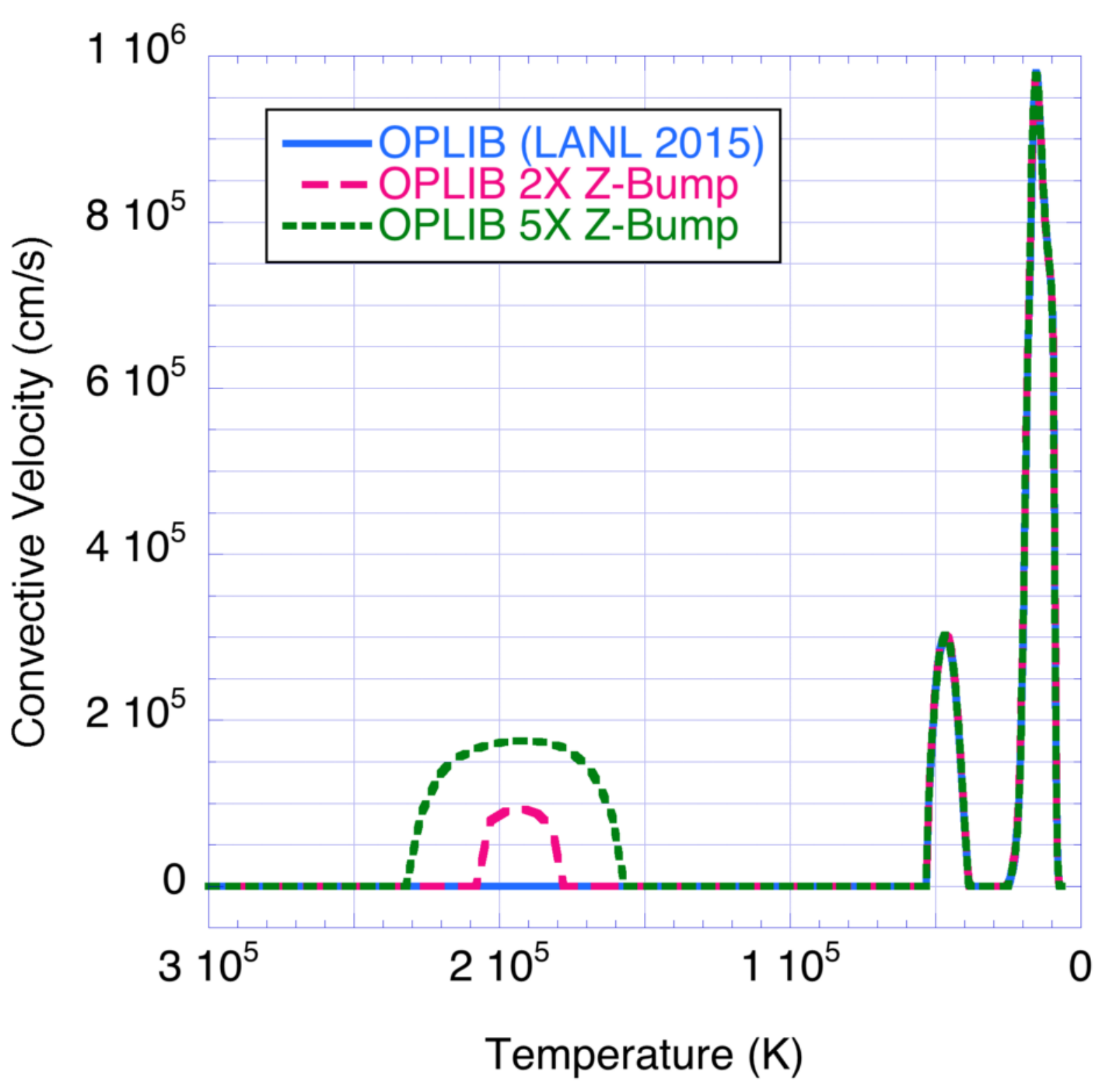}
    \includegraphics[width=0.4\textwidth]{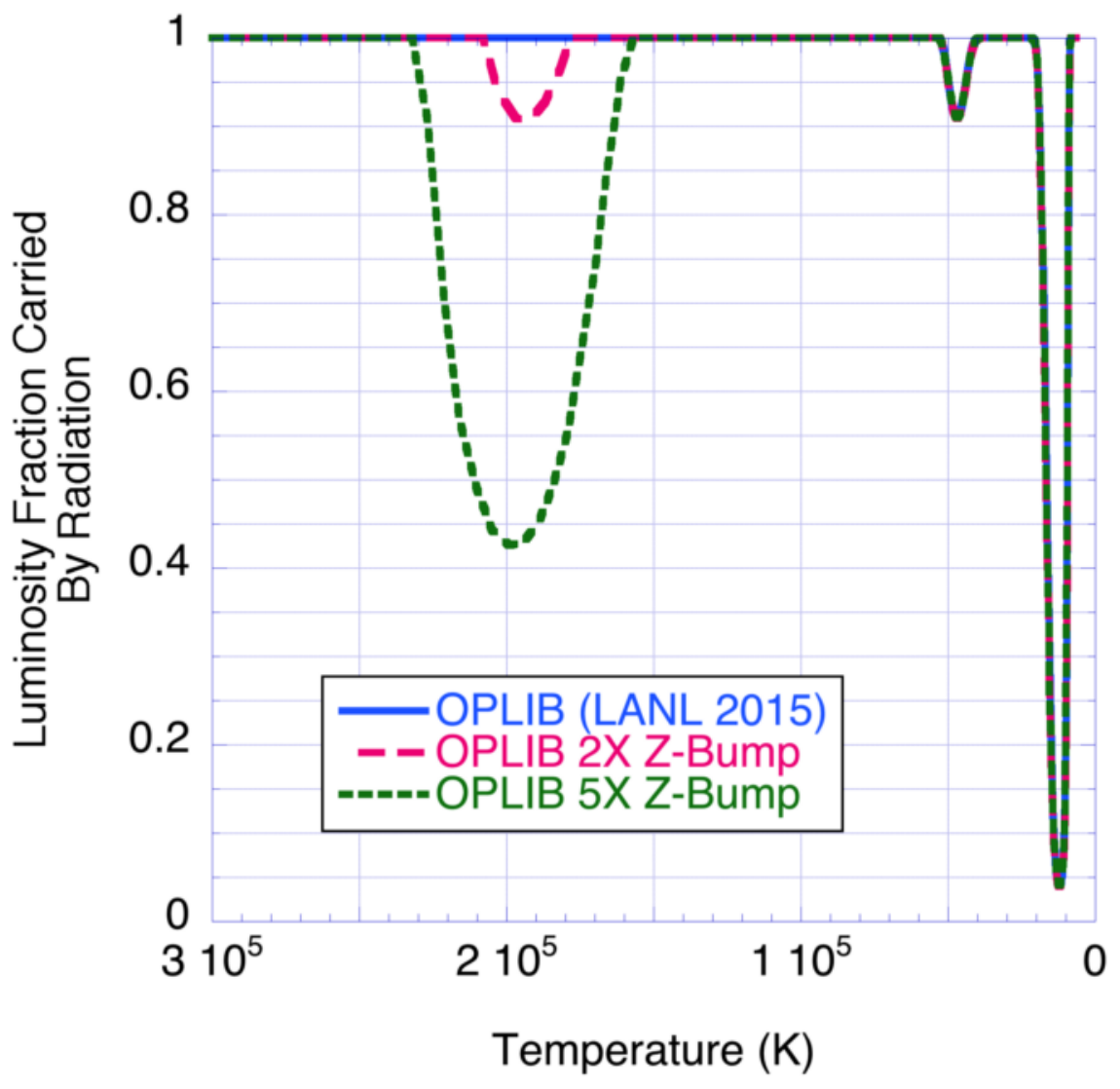}
    \caption{Convective velocity (left) and fraction of luminosity carried by radiation (right) vs.~temperature for 2 M$_\odot$ models with T$_{\rm eff}$  = 7600 K,  with and without $\times$2  and $\times$5 opacity increases in the Z-bump region.}
    \label{fig:CVLFZbump}
\end{figure}

We next explored how the evolution and $p$-mode pulsation predictions change when Z-bump opacities are increased.  Note that the $\kappa$ mechanism in the 2nd He ionization region $\sim$50,000 K drives $p$-mode pulsations in $\delta$ Sct stars, so the Z-bump at $\sim$200,000 K should not have a large effect on the $\delta$ Sct $p$ modes.   We modified the opacity by multiplying by a Gaussian function peaking at $\times$2 at 200,000 K (Fig.  \ref{fig:Opacity2xZbump}, left).  This Z-bump opacity increase has almost no effect on the 2 M$_\odot$ evolution track.  The Z-bump $\times$2 opacity increase also has almost no effect on $l$=1 $p$-mode pulsations; the growth rates are slightly lower for some modes.   However, the Z-bump opacity increase induces a small convection zone in these models near 200,000 K (Fig. \ref{fig:CVLFZbump}).  In the Z-bump region, the convective velocities are 10$^{5}$ cm/s, and convection carries only 10\% of the luminosity.

\subsection{$\times$2 He+ Ionization bump at 50,000 K}

Could neglected line broadening  be responsible for too-low stellar opacities? Turbulence and convection not taken into account in opacity simulations could broaden lines and features. We present estimates of the magnitude of this effect in Section \ref{sect:estimates} below.  It is also possible that photo-excitation processes (Stark line widths) may not be calculated properly in opacity simulations \citep{2016ApJ...824...98K}.  However, Krief et al. \cite{2016ApJ...824...98K} note that a factor of 100 increase in all line widths would be needed to account for missing solar opacity.  

\begin{figure}
\center
    \includegraphics[width=0.4\textwidth]{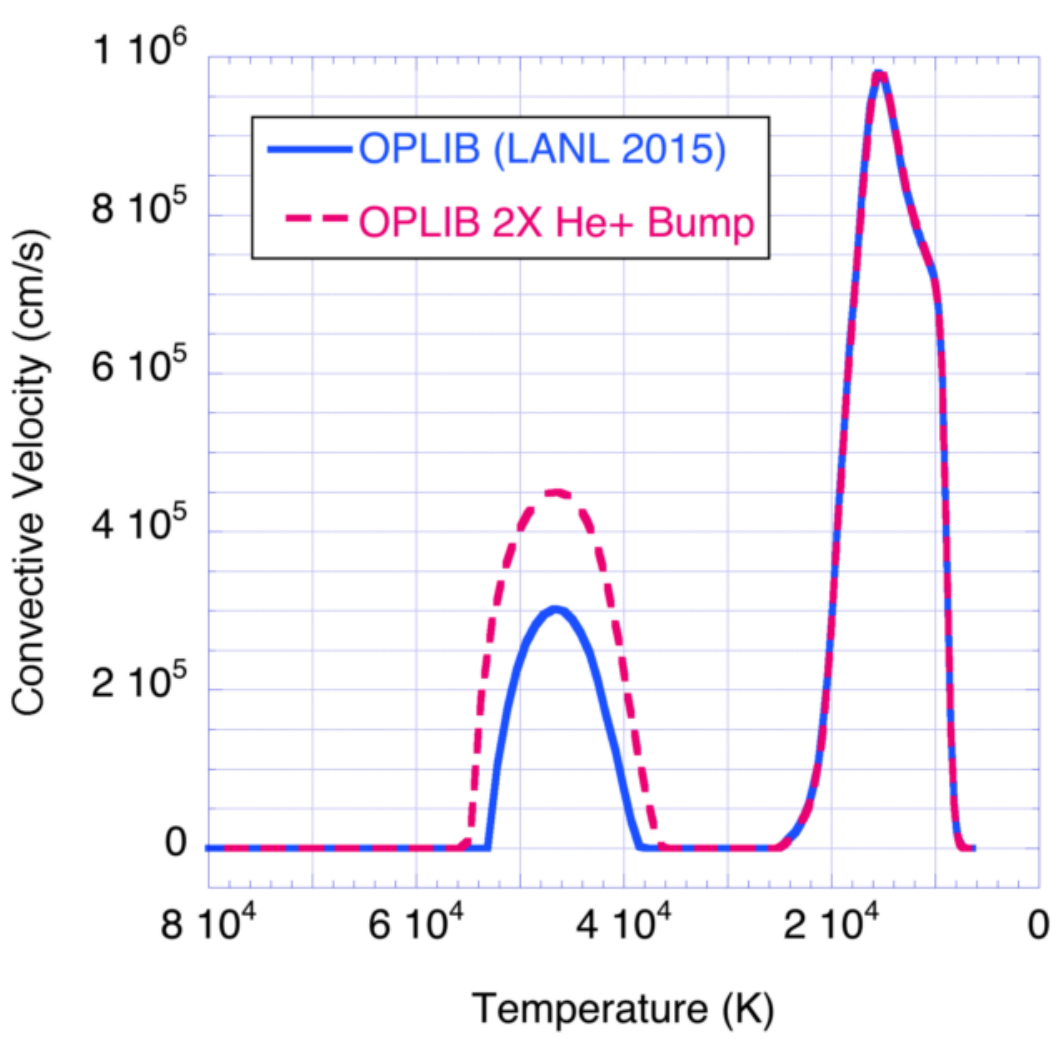}
     \includegraphics[width=0.4\textwidth]{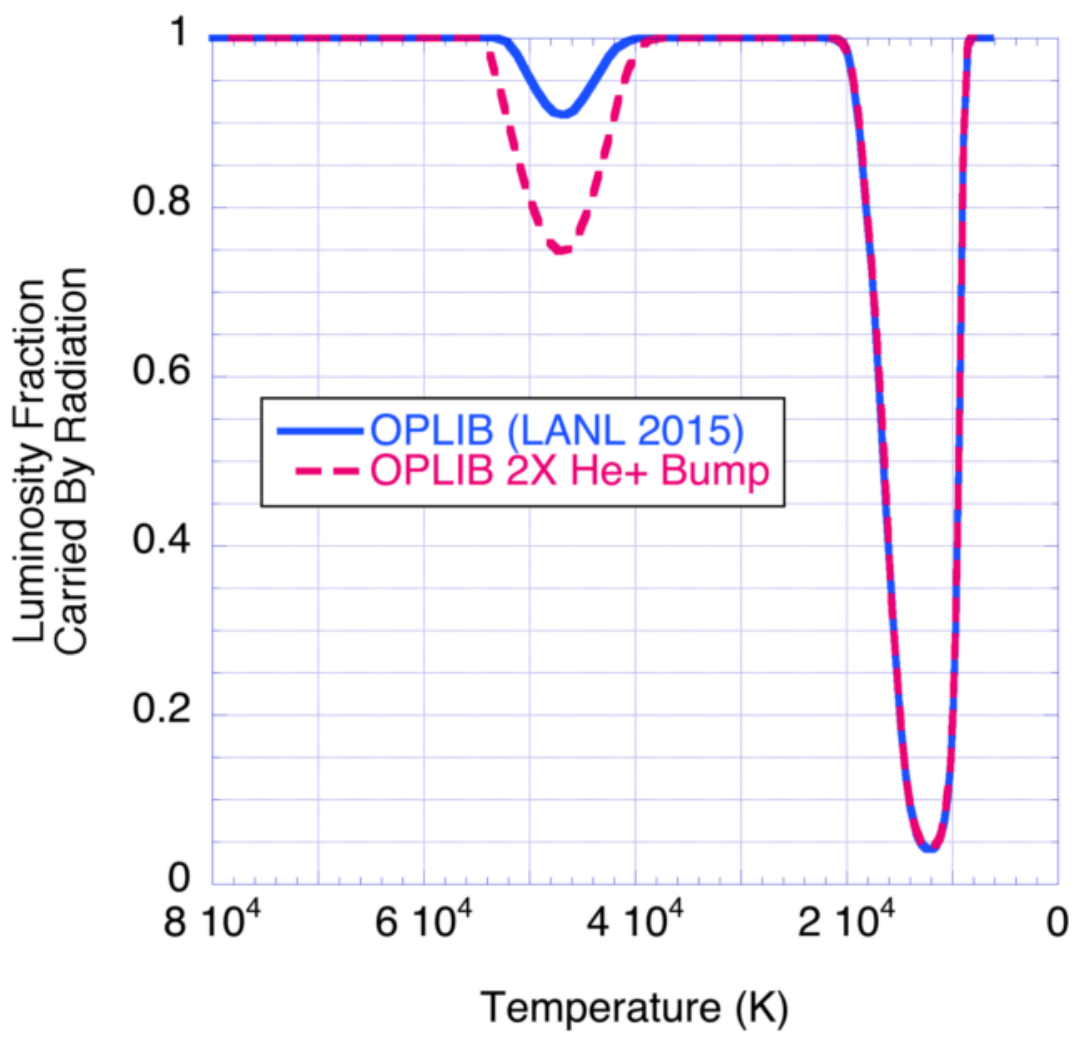}
    \caption{Convective velocity (left) and fraction of luminosity carried by radiation (right) vs.~temperature for 2 M$_\odot$ models with T$_{\rm eff}$  = 7600 K,  with and without $\times$2 opacity increase in the He+ ionization region.}
    \label{fig:CVLFHebump}
\end{figure}

What are the consequences for pulsations of A-type stars of an opacity increase in the 2nd He ionization region where $\delta$ Sct $p$ modes are driven?  This opacity bump is produced by ionization, rather than by line excitation; therefore the opacity modification may be more correctly referred to as `edge broadening' or `edge blending'. Unlike the Z-bump region around 200,000 K, the 2nd He ionization region in A-type stars responsible for $\delta$ Sct pulsations is convective/turbulent, even without opacity enhancements (Fig. \ref{fig:CVLFHebump}).

\begin{figure}
\center
    \includegraphics[width=0.4\textwidth]{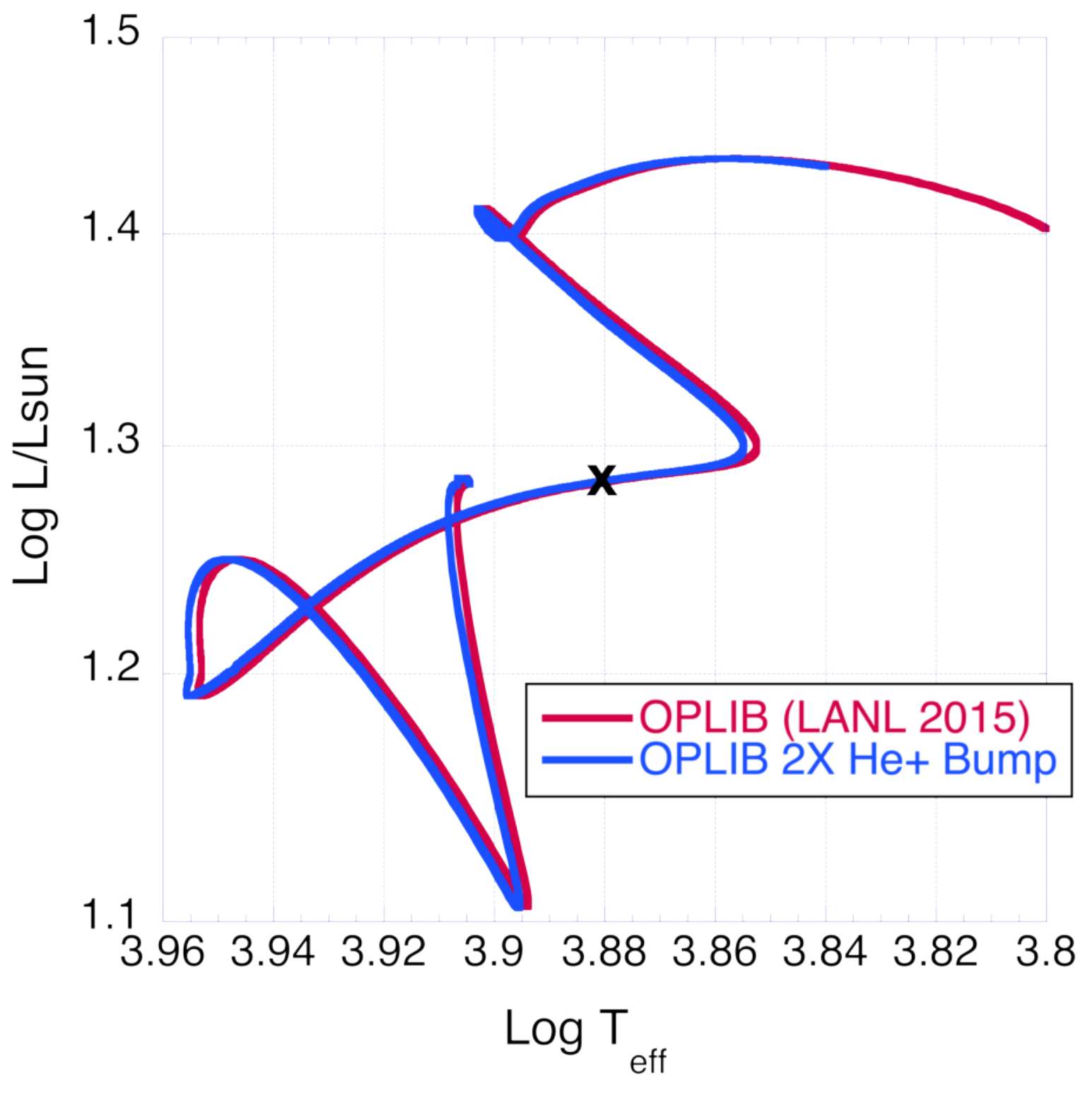}
     \includegraphics[width=0.4\textwidth]{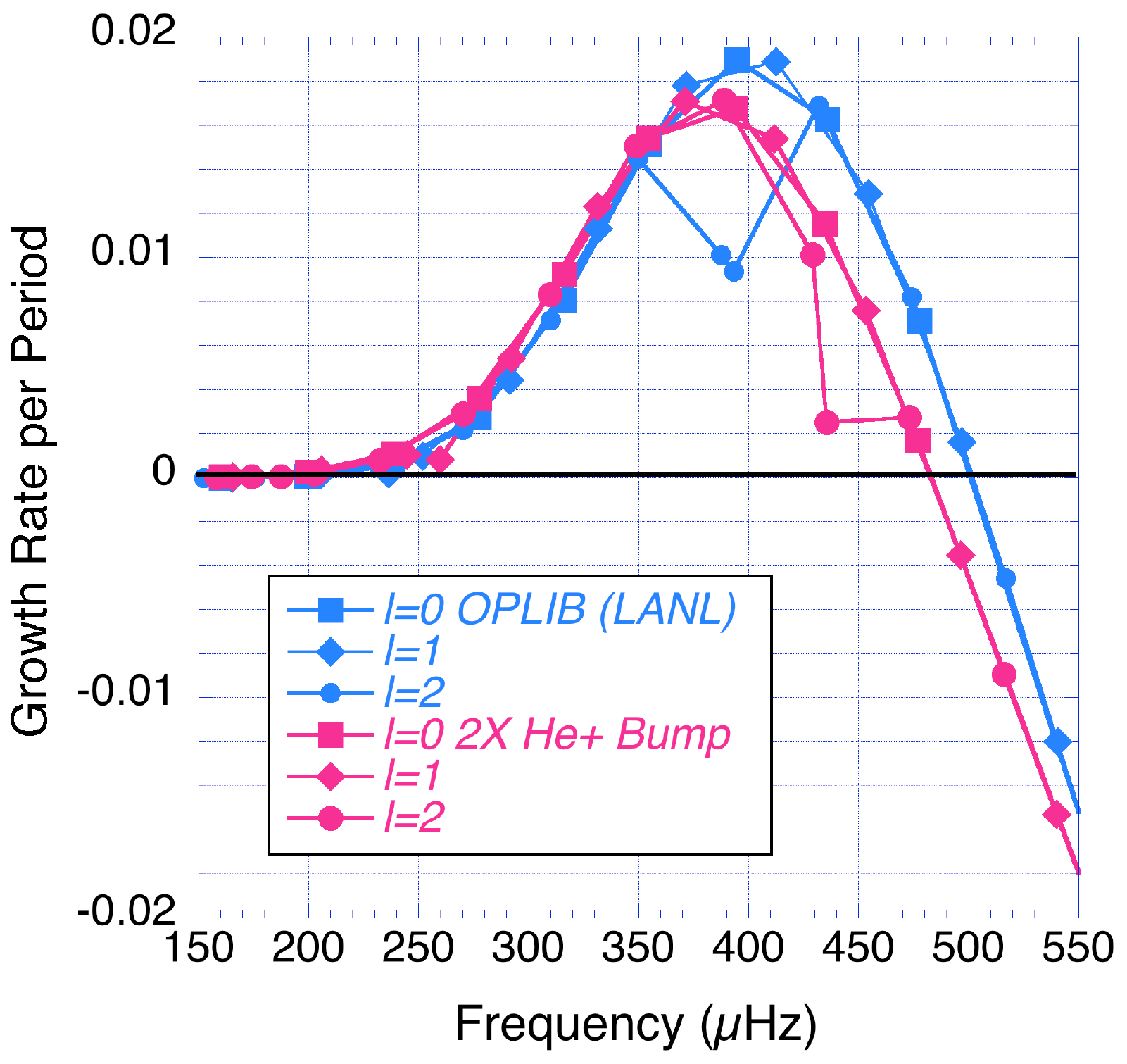}
    \caption{Evolution tracks for 2 M$_\odot$ models with and without $\times$2 opacity increase in He+ ionization region at 50,000 K (left); fractional kinetic-energy growth rate per period (dimensionless) vs.~frequency for $l$=0, 1, and 2 $p$ modes of 2 M$_\odot$ T$_{\rm eff}$  = 7600 K models with and without opacity enhancements in the He+ ionization region (left).}
    \label{fig:Evol2XHeGrowth}
\end{figure}

We multiplied the opacity by a Gaussian function peaking at $\times$2 in the 2nd He ionization region centered at 50,000 K (Fig. \ref{fig:Opacity2xZbump}, right).  We find that this He+ ionization region opacity increase has very little effect on the 2 M$_\odot$ evolution track in the H-R diagram (Fig. \ref{fig:Evol2XHeGrowth}, left).  Increased He+ opacity shifts the unstable $l$=0-2 $p$-mode frequency range to lower frequencies and also reduces the growth rates (Fig. \ref{fig:Evol2XHeGrowth}, right).  Increased He+ opacity causes more luminosity to be carried by convection, weakening the $\kappa$-effect pulsation driving.  With the opacity increase, convective velocities become larger by a factor of two, as high as 4$\times$10$^{5}$ cm/s. Convection then carries 25\% of the luminosity, instead of 10\% (Fig. \ref{fig:CVLFHebump}).

\subsection{$\times$5 Z-bump at 200,000 K}

If enhanced opacities cause a small convection zone to develop in the Z-bump region for A-type stars, convection could also cause line broadening that would further increase opacities.  The combination of diffusive settling and radiative levitation can concentrate Fe in the envelope at $\sim$200,000 K, and may also cause a Z-bump convection zone to develop for A-type stars \citep[see, e.g.,][]{2000A&A...360..603T,2012A&A...546A.100T}.

\begin{figure}
\center
    \includegraphics[width=0.4\textwidth]{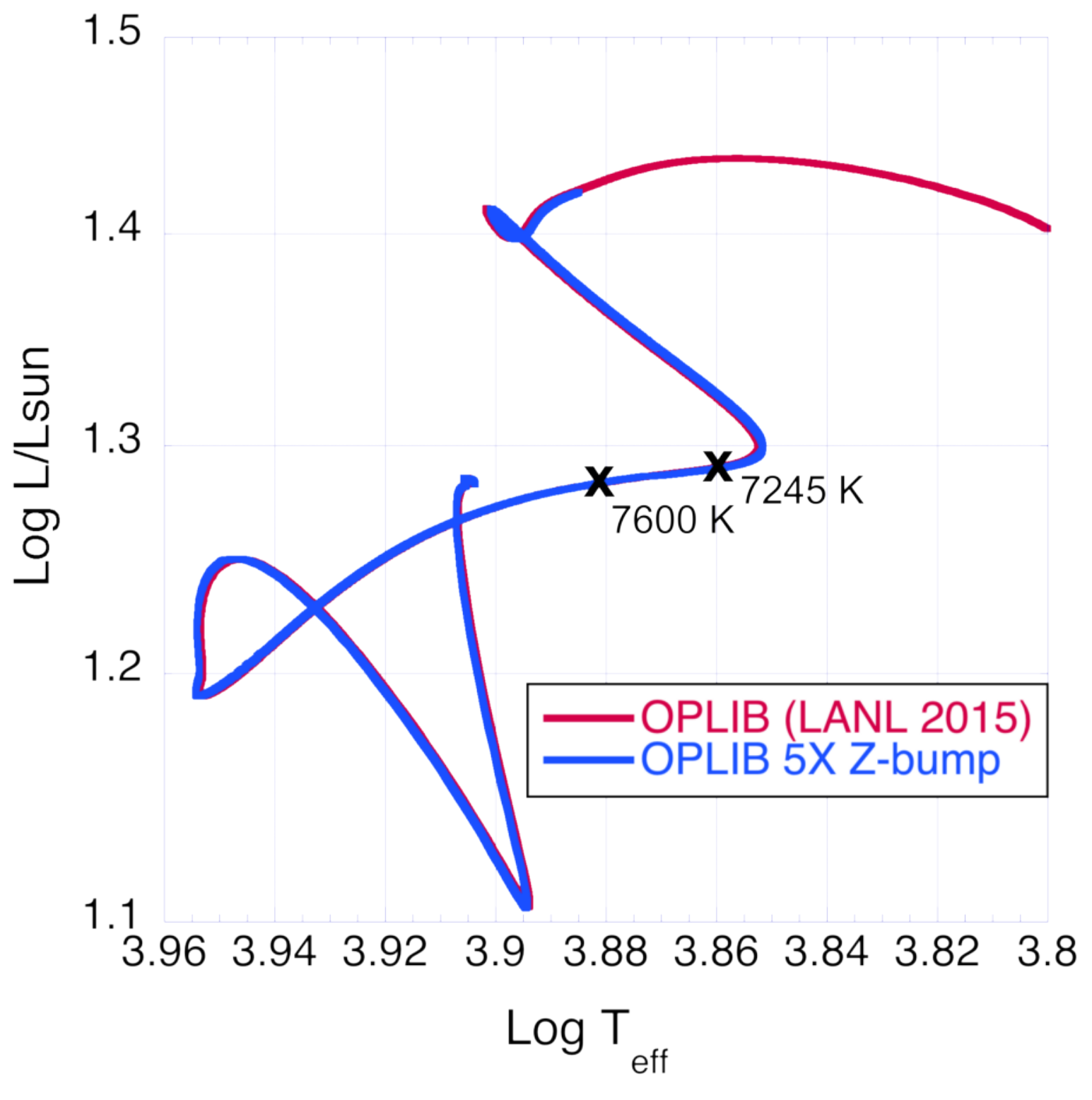}
     \includegraphics[width=0.4\textwidth]{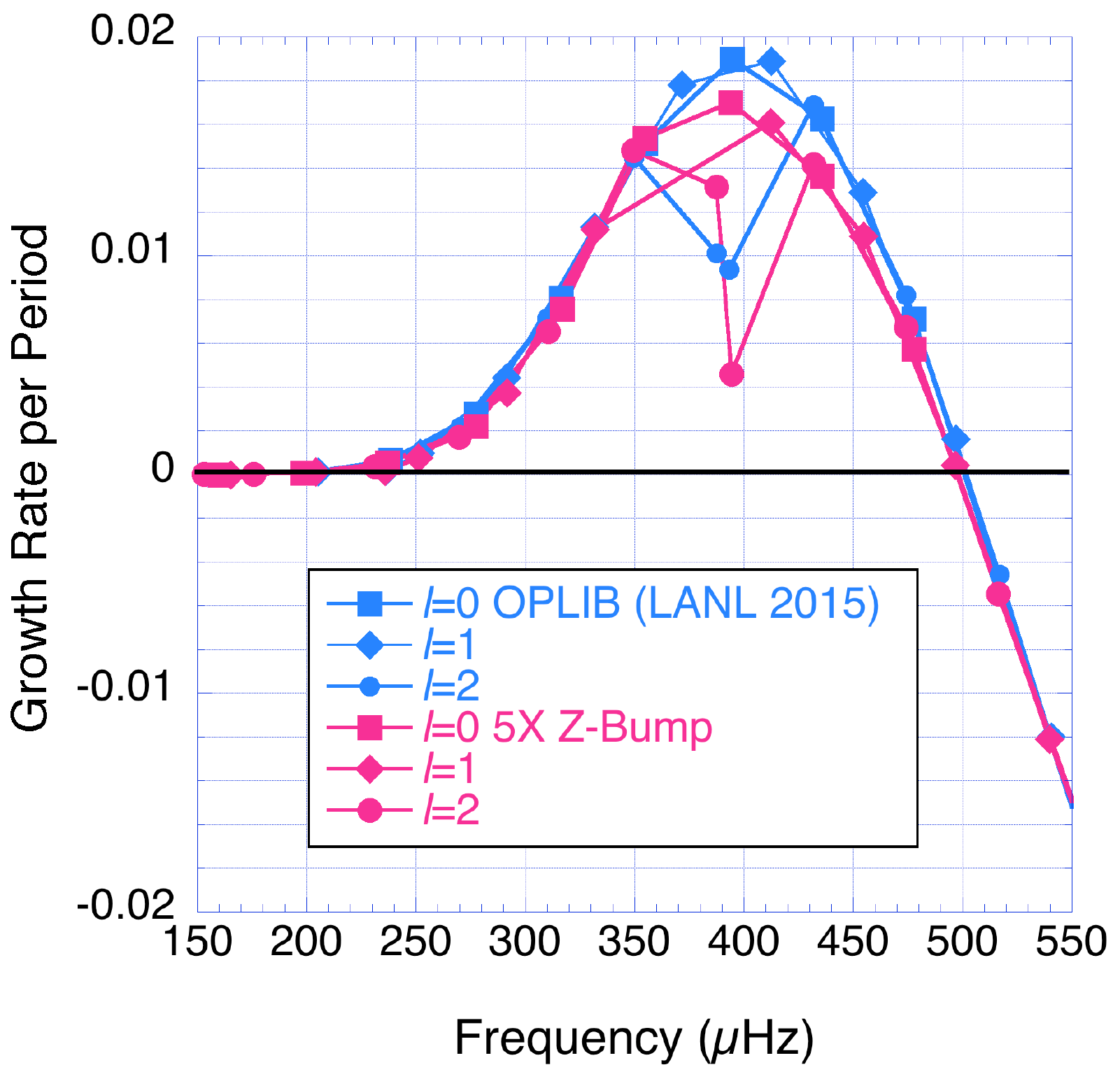}
    \caption{Evolution tracks for 2 M$_\odot$ models with and without $\times$5 opacity increase in Z-bump region (left); fractional kinetic-energy growth rate per period (dimensionless) vs.~frequency for $l$=0, 1, and 2 $p$ modes of 2 M$_\odot$ T$_{\rm eff}$  = 7600 K models with and without $\times$5 opacity enhancement in the Z-bump region (right).}
    \label{fig:EvolGrowth5xZbump}
\end{figure}

We multiplied the opacities by a Gaussian function peaking at $\times$5 for the Z-bump region centered at 200,000 K.  The  $\times$5 Z-bump opacity increase has almost no effect on the 2 M$_\odot$ evolution track in the H-R diagram (Fig. \ref{fig:EvolGrowth5xZbump}, left).  However, the $\times$5 Z-bump opacity increase induces a wide convection zone at $\sim$200,000 K, and convective velocities $\sim$2 $\times$10$^{5}$ cm/s, with convection carrying almost 60\% of the luminosity (Fig. \ref{fig:CVLFZbump}).  The Z-bump $\times$5 opacity increase has almost no effect on $l$=0-2 $p$-mode pulsations; growth rates become slightly lower for some modes (Fig. \ref{fig:EvolGrowth5xZbump}, right).

\begin{figure}
\center
    \includegraphics[width=0.4\textwidth]{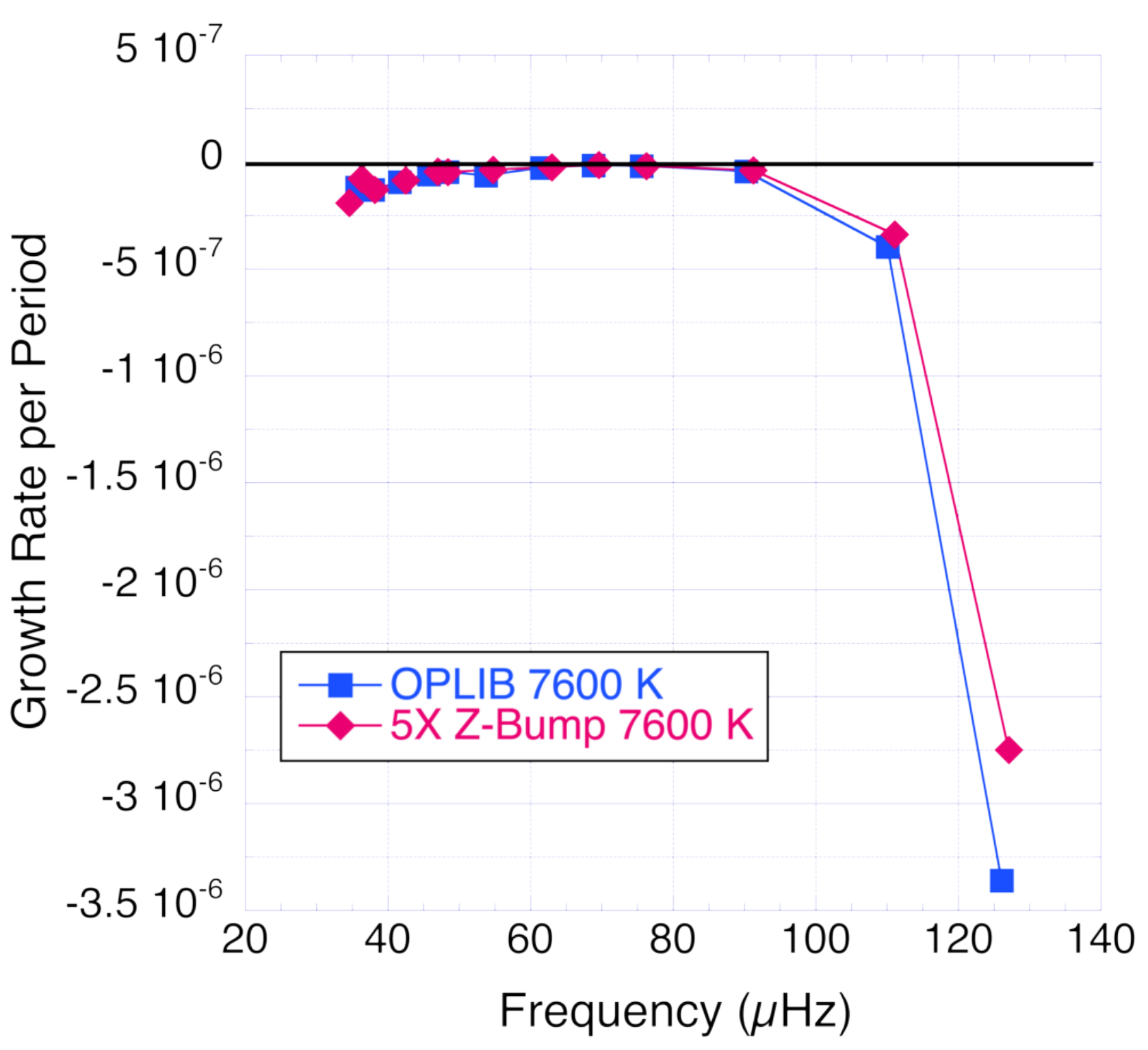}
     \includegraphics[width=0.4\textwidth]{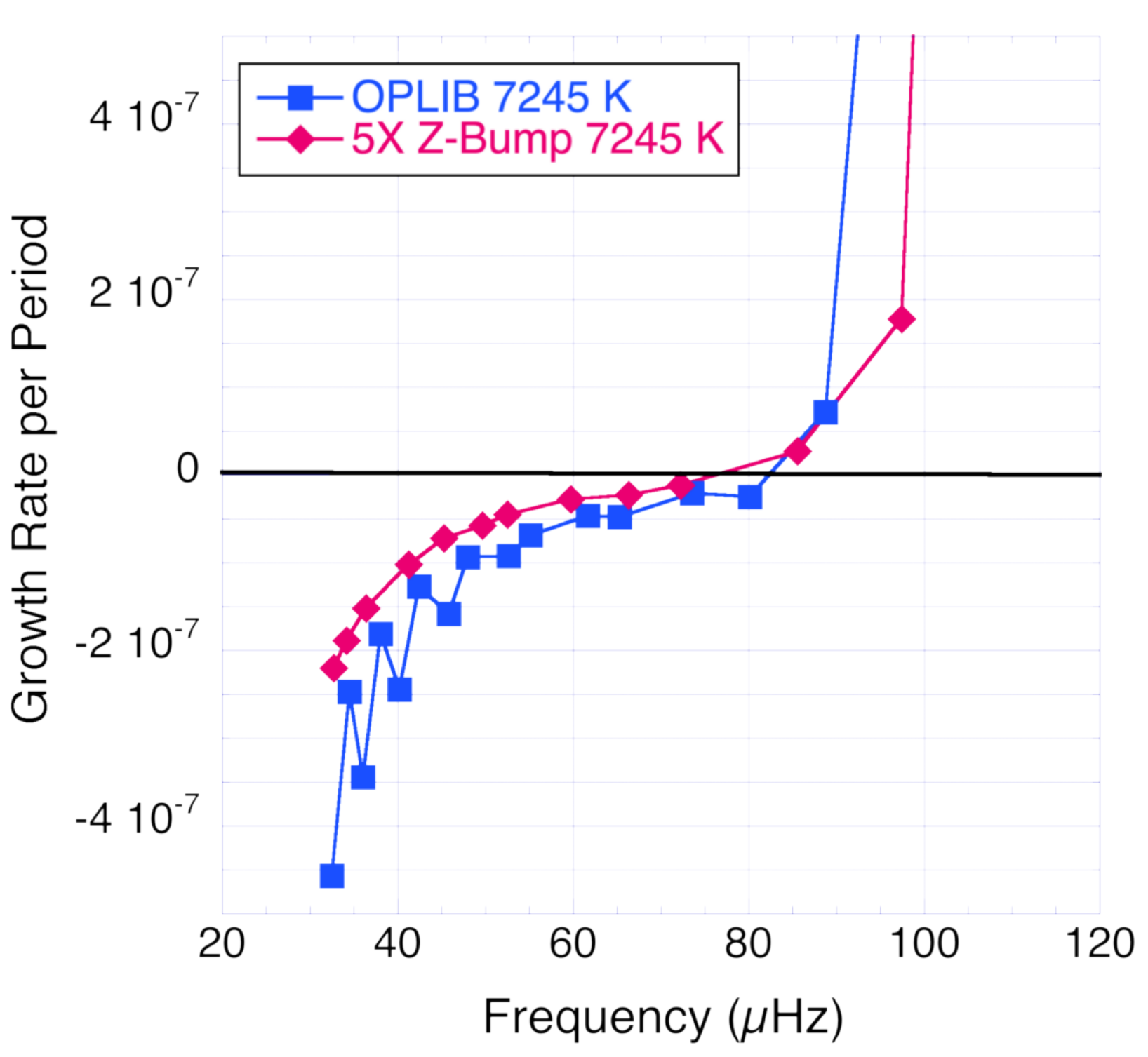}
    \caption{Fractional kinetic-energy growth rate per period (dimensionless) vs.~frequency for $l$=2 $g$ modes for models with and without $\times$5 Z-bump for T$_{\rm eff}$  = 7600 K (left) and 7245 K (right).}
    \label{fig:Growthgmodes5XZ}
\end{figure}

Long time-series high-precision photometry has revealed low-frequency $g$-mode pulsations among many A-type and $\delta$ Scuti variables that were not predicted by theoretical models \citep{2010AN....331..989G,2011A&A...534A.125U,2011MNRAS.417..591B}.  $g$-mode pulsations are observed for cooler late-A to F-type $\gamma$ Doradus variable stars, and explained by the convective-blocking mechanism \citep{2000ApJ...542L..57G} operating at the base of the deeper convective envelope that is not present in early A-type stars.  We were hoping that the new convection zone induced by opacity enhancements in the Z-bump region might cause $g$-mode pulsation driving via the convective-blocking mechanism.  However, we find that for 2 M$_\odot$T$_{\rm eff}$ = 7600 K models, the $\times$5 Z-bump enhancement does not affect low-order $l$=1 or 2 $g$-mode stability; the low-order $g$ modes have slightly less negative growth rates, but are still stable (Fig. \ref{fig:Growthgmodes5XZ}, left).  One or two of the highest frequency $l$=2 $g$ modes become unstable for cooler (T$_{\rm eff}$  = 7245 K) models, even without an opacity enhancement (Fig. \ref{fig:Growthgmodes5XZ}, right).  The growth rates become lower with the $\times$5 Z-bump increase, likely because these modes are actually driven by the $\kappa$ effect (and not the convective-blocking mechanism) that is weakened by convection.

We also calculated frequencies for a 2 M$_\odot$ model closer to the blue edge of the instability region at T$_{\rm eff}$  = 8318 K.   For this T$_{\rm eff}$, there are only three unstable $p$ modes at frequencies 420-580 $\mu$Hz for each degree $l$=0, 1, and 2 with the $\times$5 Z-bump enhancement.  There are the same number of unstable modes for the comparable model with no Z-bump enhancement, with the exception of a fourth unstable mode for $l$=0.  Again, we do not find any unstable $g$ modes.  Note that \cite{2015MNRAS.452.3073B} conclude that it is possible to create $g$-mode instability with a Z-bump enhancement of at least $\times$3, based on results from their more extensive parameter study.

Daszy{\'n}ska-Daszkiewicz et al. \citep{2017MNRAS.466.2284D} investigate enhanced opacities for B-type stars at somewhat higher interior temperature, log T = 5.46, or T= 288,400 K, corresponding to the region of maximum contribution of nickel to the opacity.  Because lower-frequency $g$-mode pulsations that have also been seen in many A-type stars are driven most effectively at temperature of about 350,000 K in the cooler, less massive $\gamma$ Dor variables \citep{2000ApJ...542L..57G}, we also explored a $\times$2 opacity bump at 350,000 K.  However, introducing additional opacity at this temperature, unlike for the 200,000 K Z-bump region, did not induce a new convective region, or significantly change the evolution or $p$-mode frequencies, or result in driving of $g$-mode pulsations.

\begin{figure}
\center
    \includegraphics[width=0.4\textwidth]{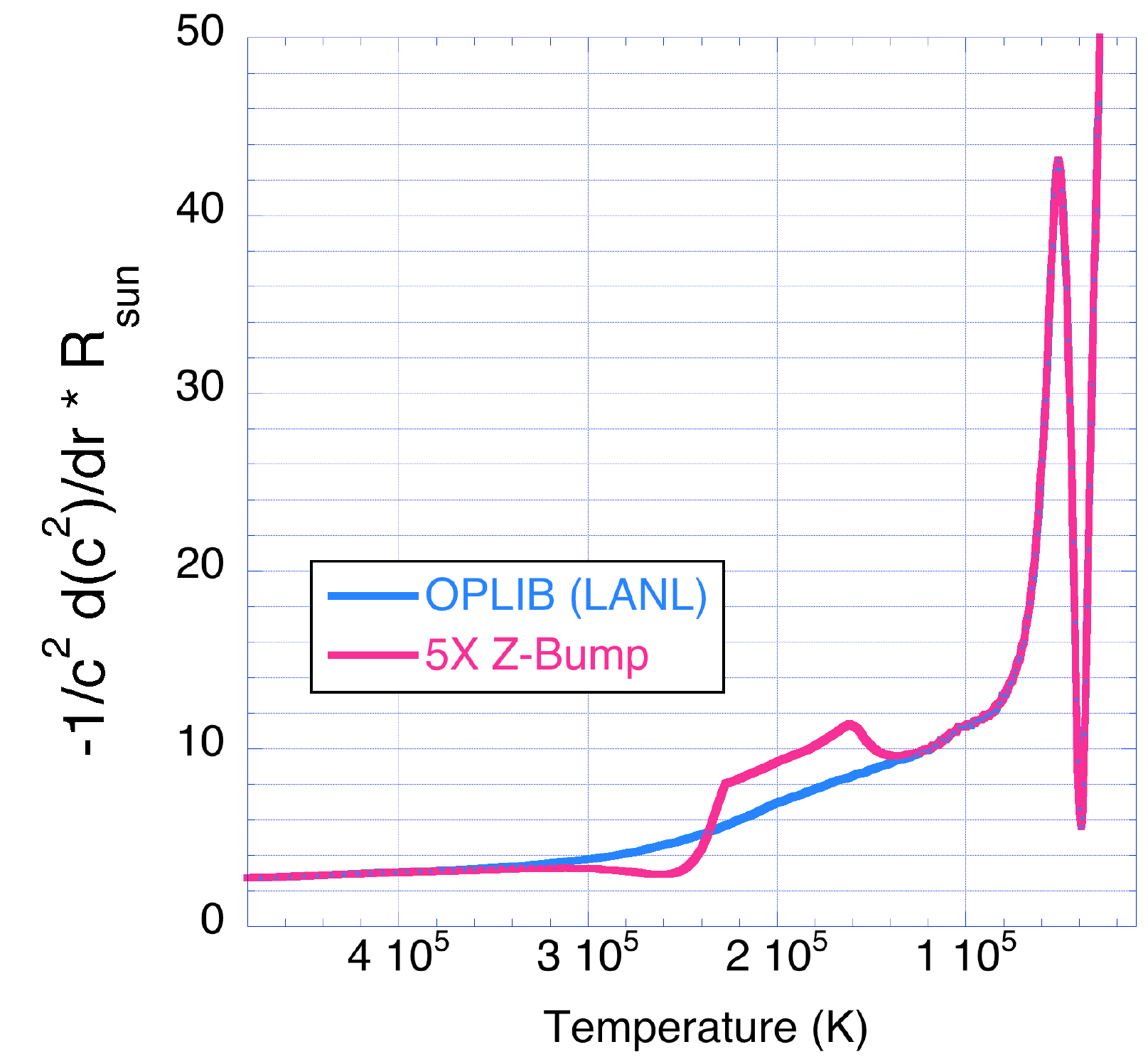}
       \caption{Radial derivative of squared sound speed vs. temperature for 2 M$_\odot$ T$_{\rm eff}$  = 7600 K models with and without $\times$5 opacity enhancement in the Z-bump region at 200,000 K.  The change in sound-speed derivative induced by the development of a convective region for the $\times$5 Z-bump model may potentially be observable applying sound-speed inversion techniques.}
    \label{fig:SoundSpeed}
\end{figure}

Would higher opacities that extend a convective region or induce a new convective region have observable signatures?  Deal  et al. \cite{2016A&A...589A.140D, 2017sf2a.conf...31D} discuss the effects of molecular-weight gradient instabilities and ``fingering convection'' induced by localized element accumulation via diffusive settling and radiative levitation on the evolution and surface abundances of $\delta$ Sct stars.  They find that surface abundances of C, N, O, Ne, Mg, Ca and Fe can be altered if fingering convection connects the surface convective layers to deeper layers.  A transition from a radiative to convective region at the Z-bump location could also have a signature in asteroseismic inversions for example, on the inferred sound-speed gradient.   Fig. \ref{fig:SoundSpeed} compares the radial derivative of the squared sound speed vs. temperature for models with and without the $\times$5 Z-bump enhancement.  High precision observations of $\delta$ Sct variables showing many $p$ modes of several angular degrees may be amenable to inversion techniques that could detect changes in the sound-speed gradient. 

\subsection{$\times$1.5 opacity bump at 2.6 million K}

\begin{figure}
\center
    \includegraphics[width=0.4\textwidth]{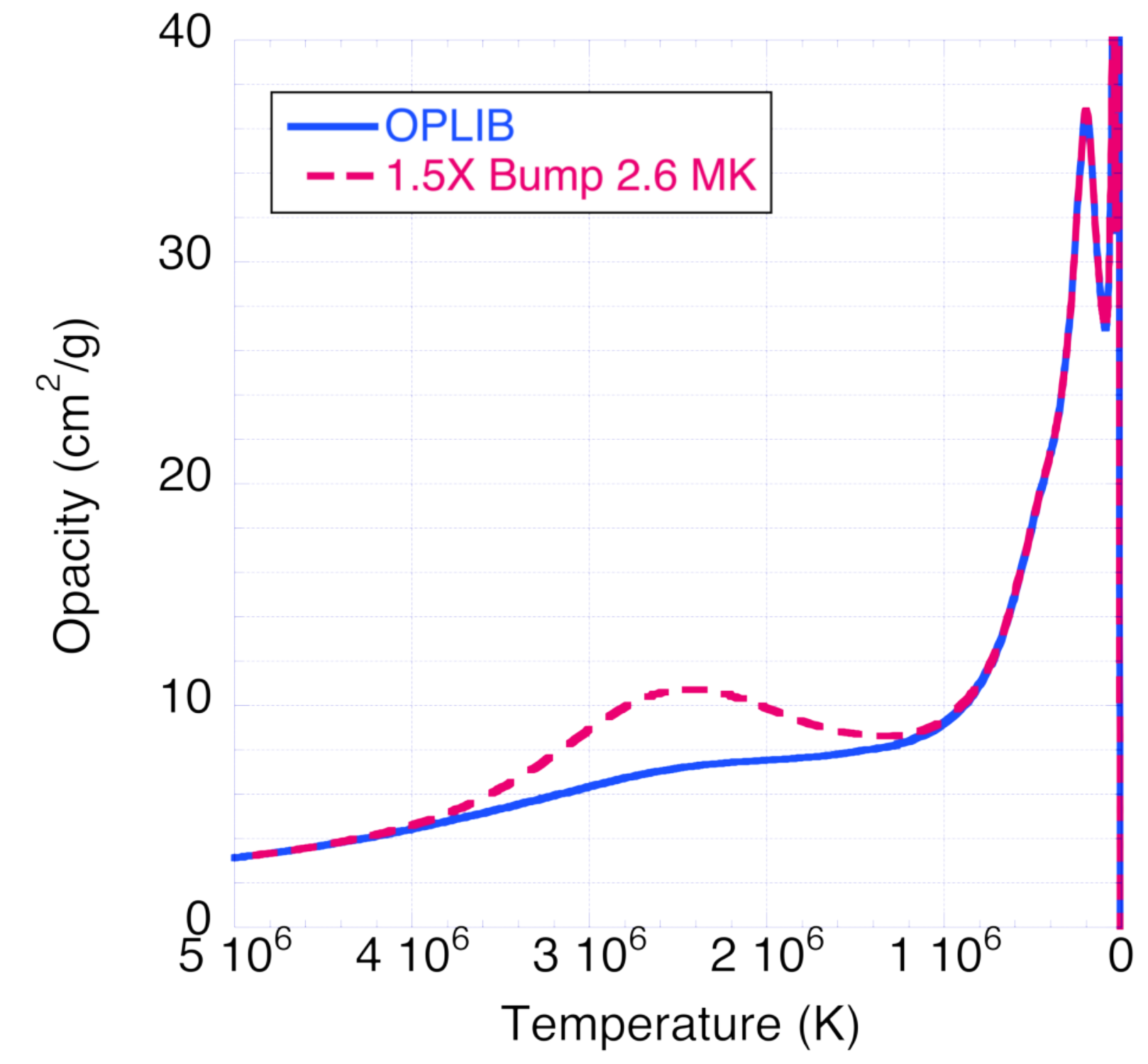}
     \includegraphics[width=0.4\textwidth]{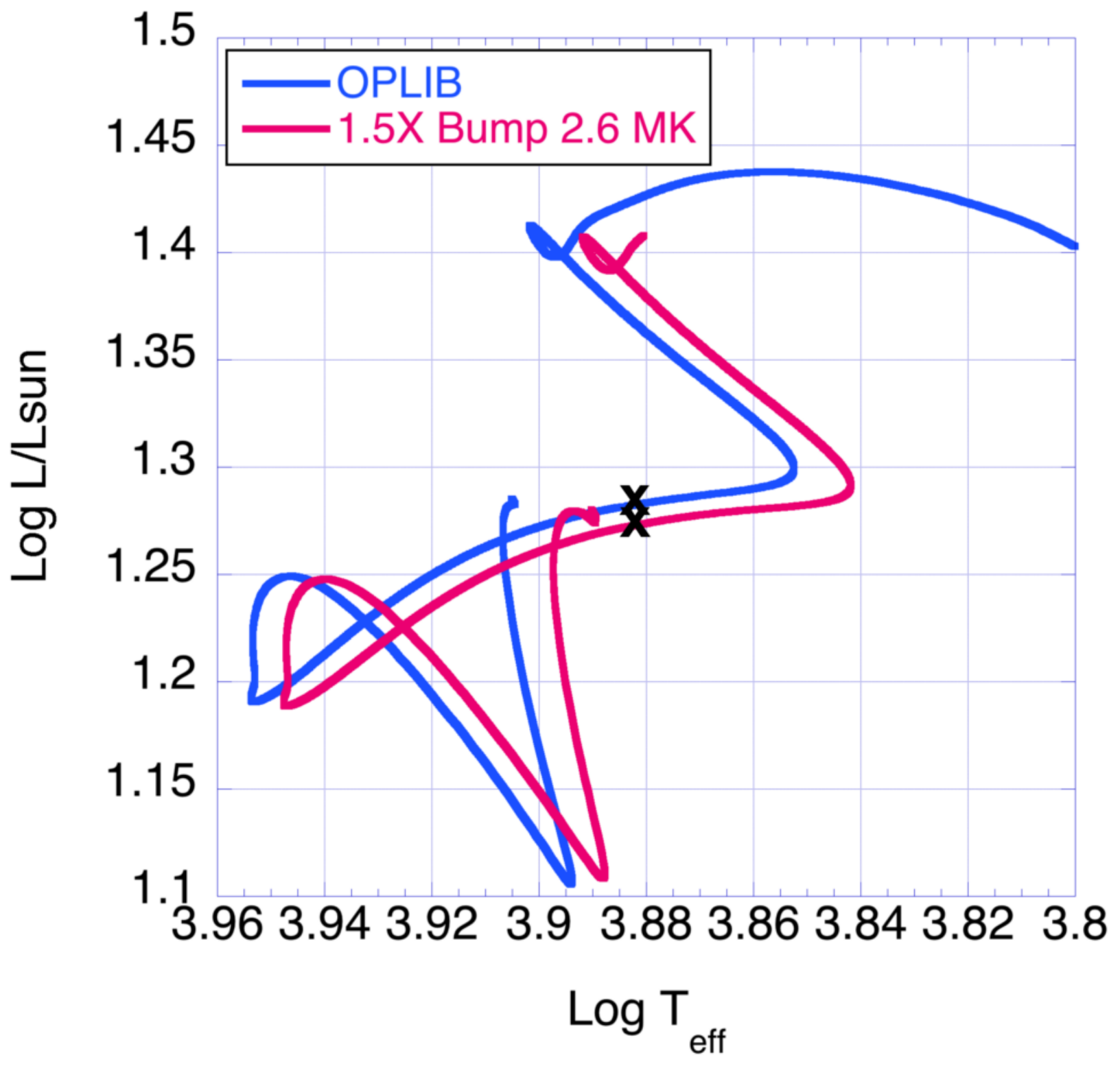}
    \caption{Opacity vs. stellar interior temperature for 2 M$_\odot$ models with T$_{\rm eff}$  = 7600 K, showing opacity multiplied by Gaussian function centered at 2.6 million K with $\times$1.5 peak (left); evolution track in H-R diagram for 2 M$_\odot$ models with and without opacity modification (right). The $\bf{X}$s mark the T$_{\rm eff}$  = 7600 K model positions on the evolution tracks.}
    \label{fig:OpacEvol2p6MK}
\end{figure}

\begin{figure}
\center
    \includegraphics[width=0.4\textwidth]{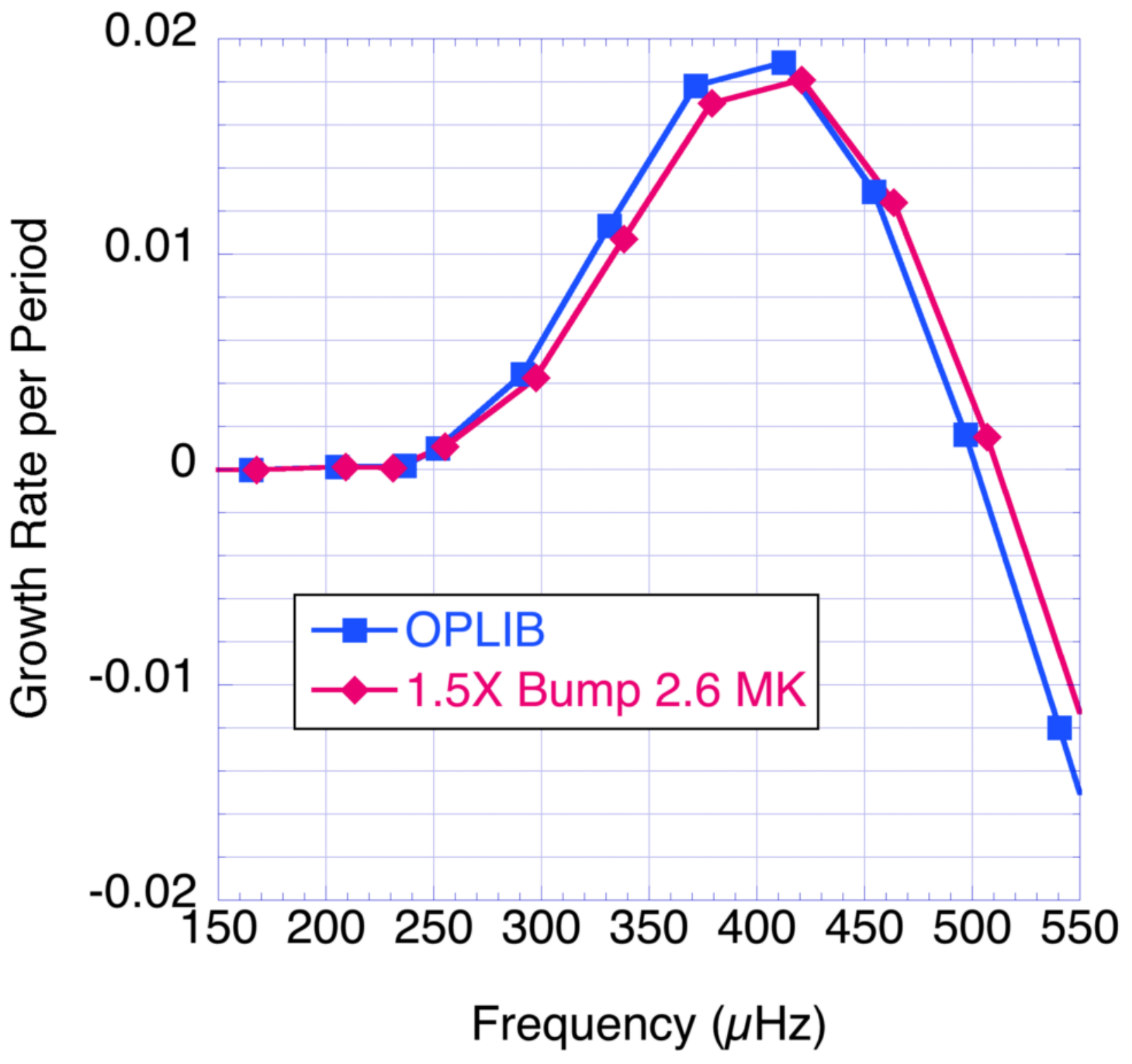}
     \includegraphics[width=0.4\textwidth]{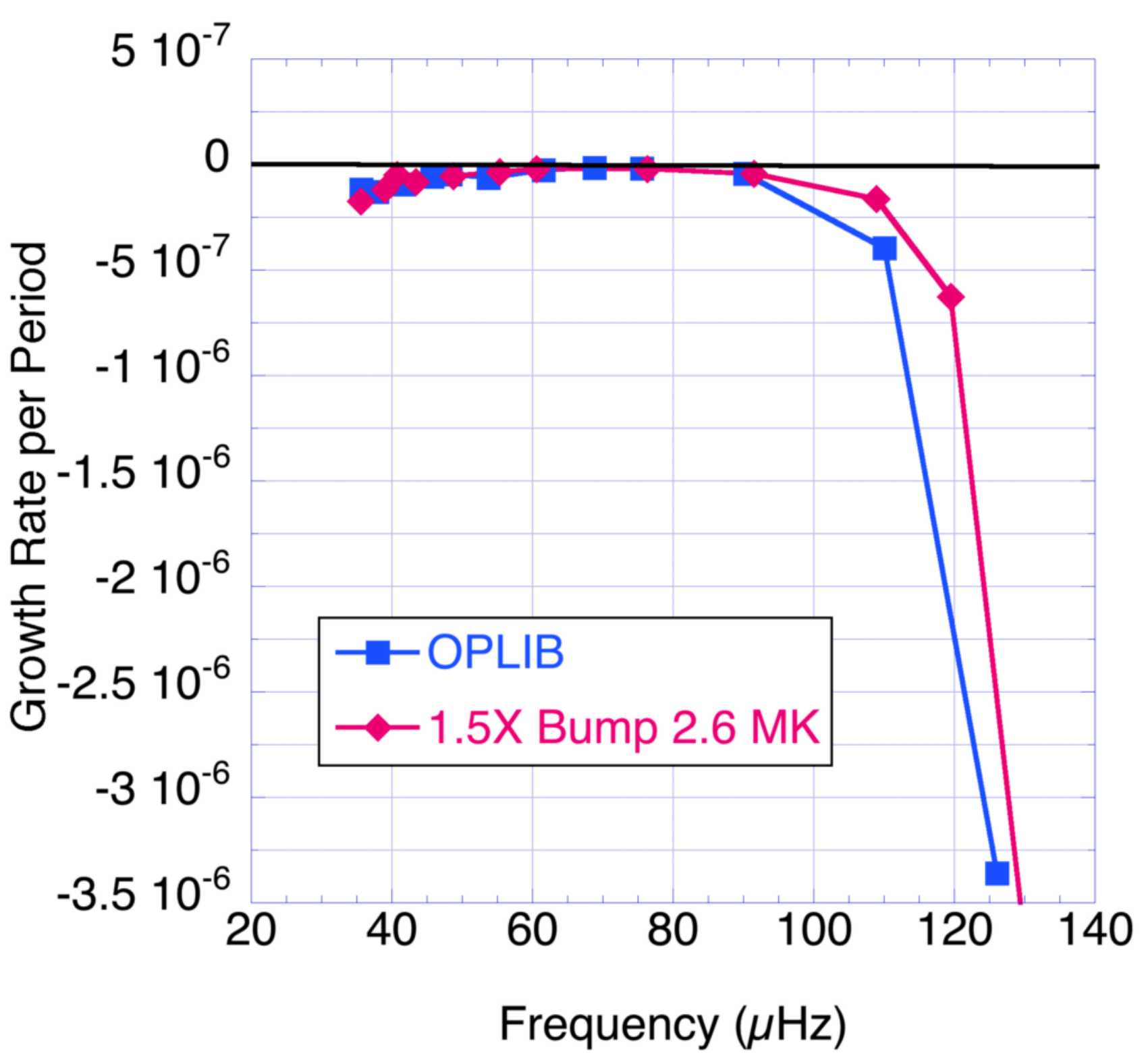}
    \caption{Fractional kinetic-energy growth rate per period (dimensionless) vs.~frequency for $l$=1 $p$ modes (left) and $l$=2 $g$ modes (right) for T$_{\rm eff}$  = 7600 K models with and without $\times$1.5 opacity enhancement at 2.6 million K.}
    \label{fig:Growth2p6MK}
\end{figure}

In addition to an opacity enhancement at the Z-bump, Zdravkov and Pamyatnykh \citep{2009AIPC.1170..388Z} advocate a 16\% opacity increase near a deeper bump around 2 million K to improve agreement with pulsation observations for hybrid $\beta$ Cep/SPB star $\gamma$ Peg.  In our cooler A-type stars, this deeper bump occurs around 2.6 million K.  We introduced a 50\% opacity enhancement peaking at 2.6 million K (Fig. \ref{fig:OpacEvol2p6MK}, left) for the 2 M$_\odot$ evolution models.  This relatively deep opacity enhancement has a more substantial effect on the 2 M$_\odot$ evolution track in the H-R diagram, causing the model to evolve at cooler temperatures and slightly lower luminosity (Fig. \ref{fig:OpacEvol2p6MK}, right).  However, the effects on $l$=1 $p$-mode pulsations (Fig. \ref{fig:Growth2p6MK}, left) and $l$=2 $g$-mode pulsations (Fig. \ref{fig:Growth2p6MK}, right) for T$_{\rm eff}$  = 7600 K models are again very small.

\subsection{$\times$2 bump at 13,000 K in H and 1st He ionization region}

It is also possible that turbulence could increase opacity in the H and 1st He ionization region around 13,000 K.   In this region there is a high convective velocity, $\sim$2 $\times$ 10$^{6}$ cm/s, and convection carries over 90\% of the luminosity (Fig. \ref{fig:CVLFHebump}).  Turbulent pressure (and its perturbation) in the H-ionization zone has been proposed as a mechanism to drive coherent modes in A-type stars \citep{2014ApJ...796..118A}, as seen in, e.g., HD 187547 observed by the {\it Kepler} spacecraft.  Stochastically excited solar-like $p$ modes were also anticipated for $\delta$ Sct stars \citep{2002A&A...395..563S}.

\begin{figure}
\center
    \includegraphics[width=0.4\textwidth]{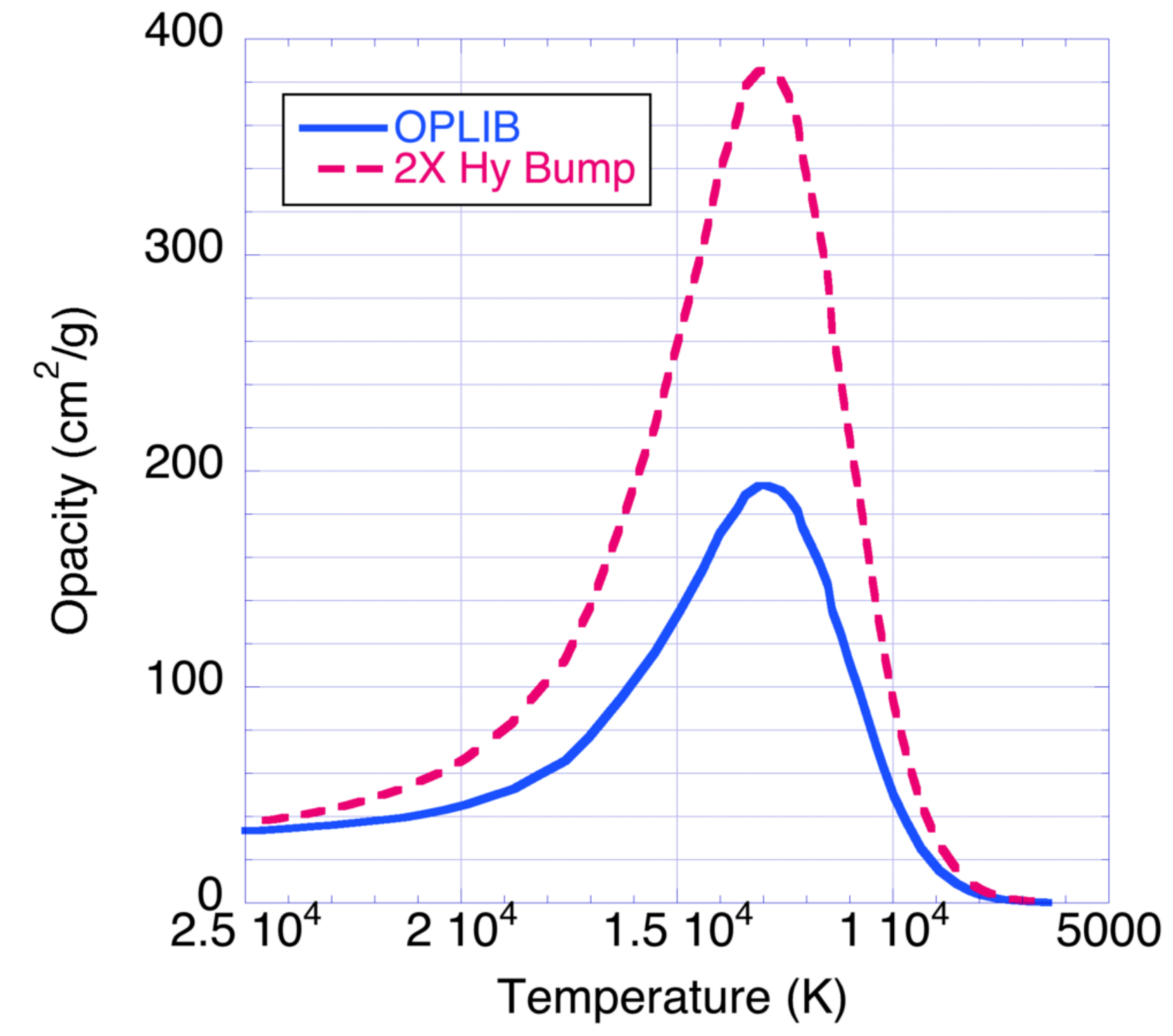}
     \includegraphics[width=0.4\textwidth]{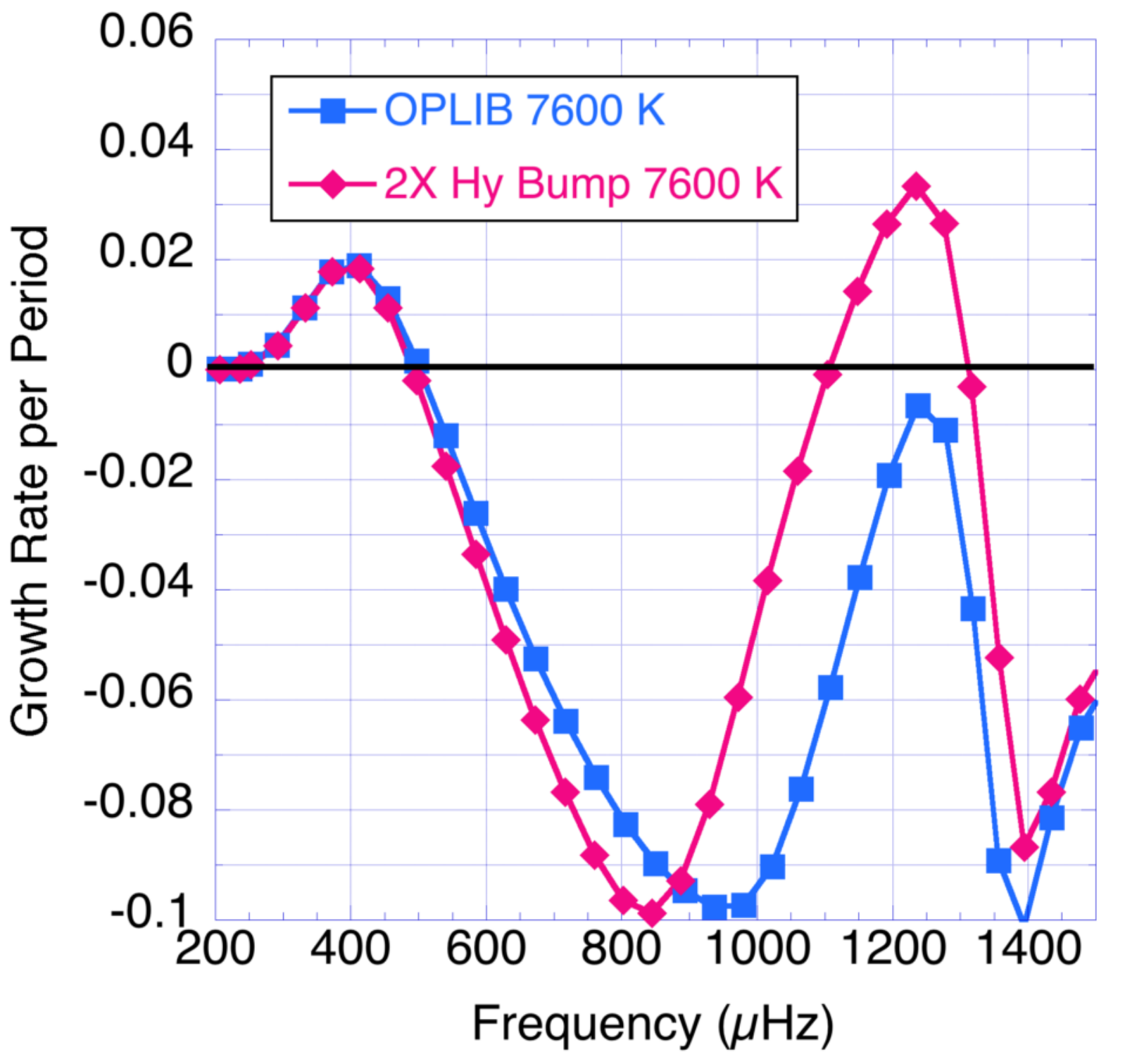}
    \caption{Opacity vs. interior temperature for 2 M$_\odot$ models with T$_{\rm eff}$  = 7600 K, showing modification of the hydrogen and 1st He ionization opacity bump at 13,000 K (left); fractional kinetic-energy growth rate per period (dimensionless) vs.~frequency for $l$=1 $p$ modes for T$_{\rm eff}$ = 7600 K models (right).}
    \label{fig:OpacityGrowth2XHybump}
\end{figure}

\begin{figure}
\center
    \includegraphics[width=0.4\textwidth]{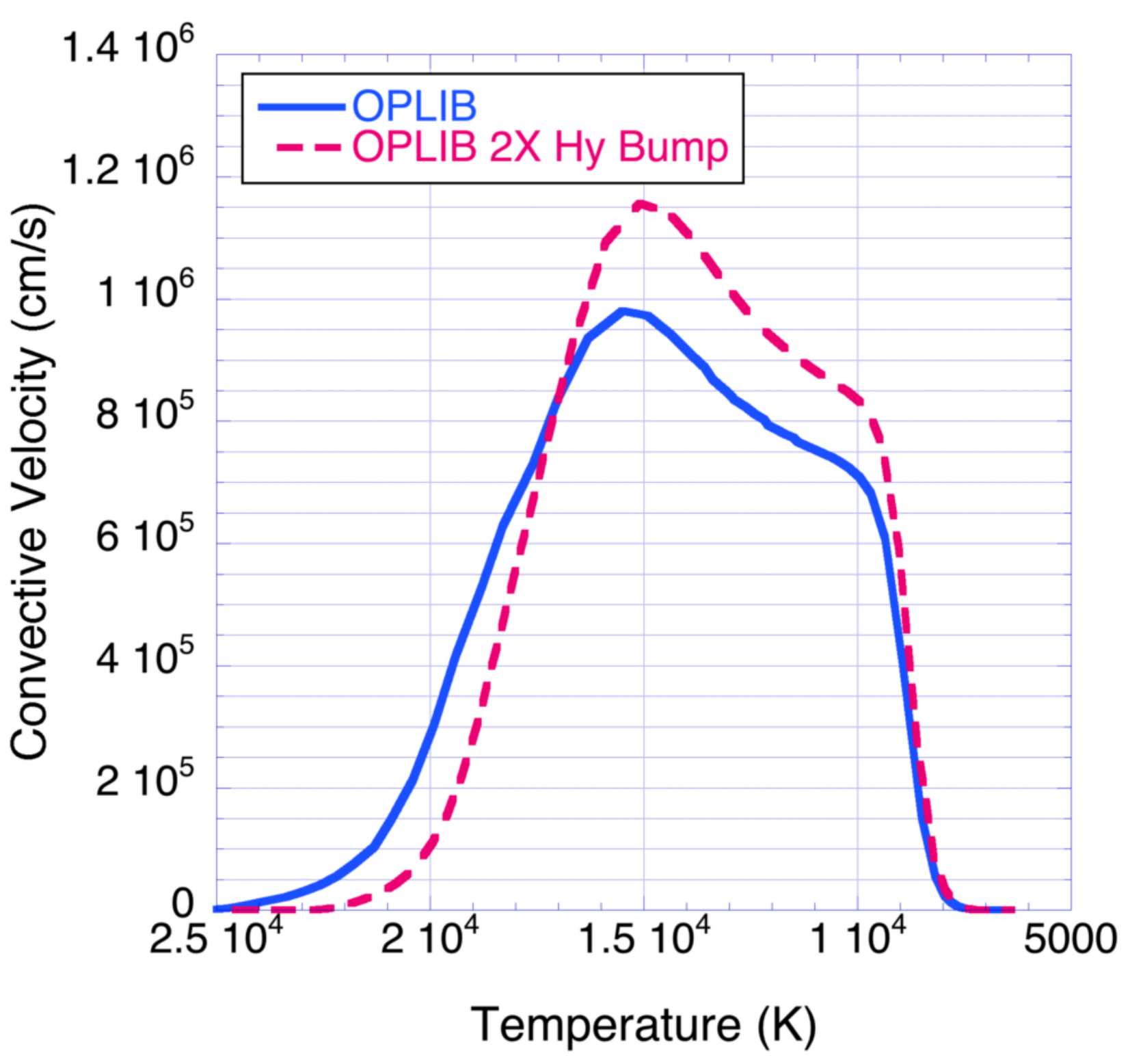}
     \includegraphics[width=0.4\textwidth]{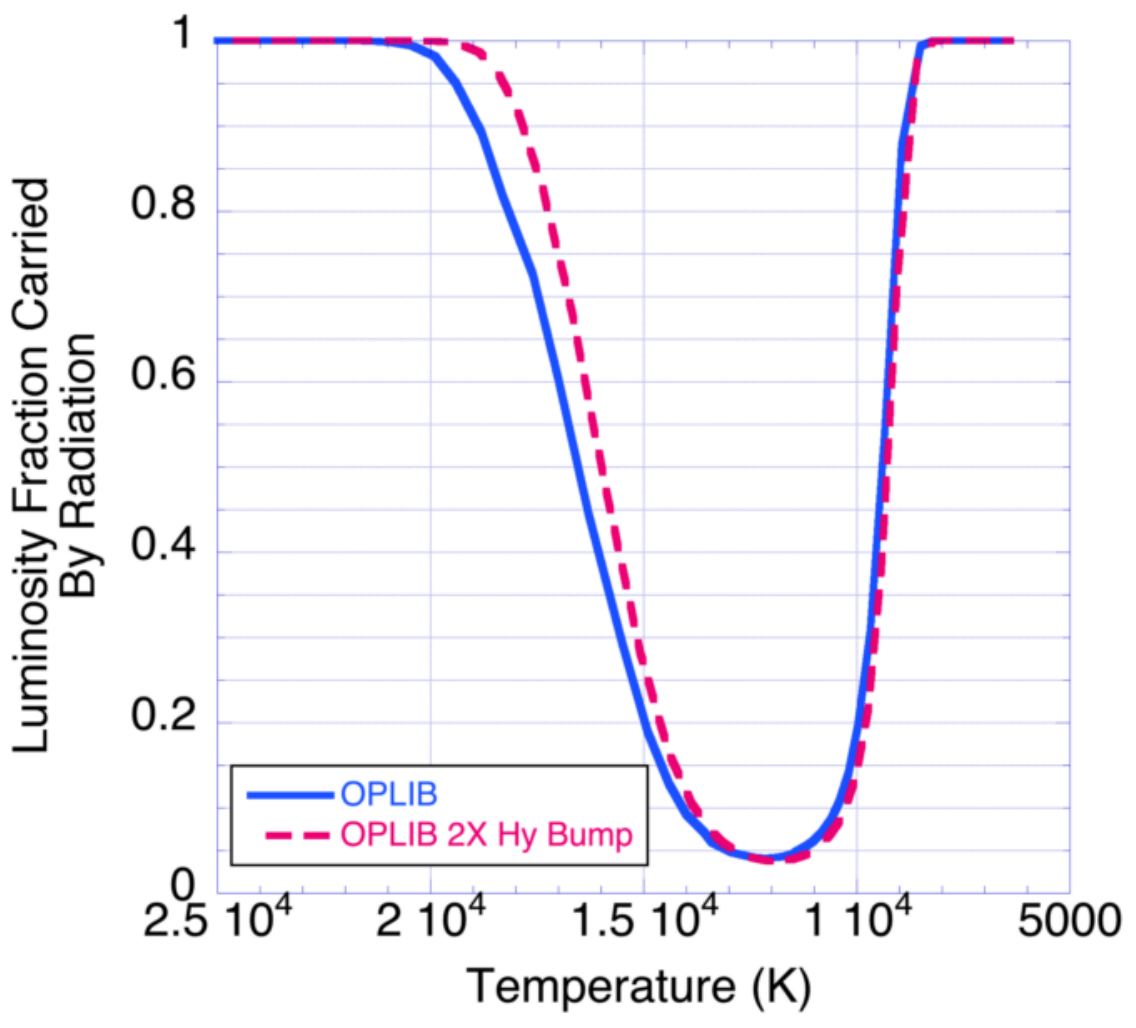}
    \caption{Convective velocity (left) and fraction of luminosity carried by radiation (right) vs.~temperature for 2 M$_\odot$ models with T$_{\rm eff}$  = 7600 K,  with and without $\times$2 opacity increase in the hydrogen and first He ionization region.}
    \label{fig:CVLFHybump}
\end{figure}

We multiplied the opacity bump in this region by a Gaussian function peaking at $\times$2 centered at 13,000 K (Fig. \ref{fig:OpacityGrowth2XHybump}, left).  This opacity enhancement has almost no effect on the evolution track in the H-R diagram.  The opacity enhancement only slightly increases the convective velocity (Fig. \ref{fig:CVLFHybump}, left) and widens the convective region (Fig. \ref{fig:CVLFHybump}, right).  However, the enhanced opacity results in predicted driving of additional $p$ modes at higher frequencies (Fig. \ref{fig:OpacityGrowth2XHybump}, right).  This pulsation driving prediction is not reliable because of the frozen-in convection treatment of our pulsation modeling code.  However, the appearance of additional $p$ modes suggests that improved treatments of the surface layers, including treatments of turbulent pressure and time-dependent convection, along with opacities, may help to explain unexpected pulsation modes found in A-type stars.  Observations and characterization of higher-frequency modes in $\delta$ Sct stars would be useful to help constrain the many physical processes in the near-surface layers of these stars, including diffusive settling and levitation, development of peculiar chemical abundances, flares and magnetic activity, extent of convection zones and convective overshoot, turbulent pressure, opacities, mechanisms for mode selection and regulating mode amplitudes, and would provide additional asteroseismic constraints for inversions.  

\section{Summary of opacity and model parameters}

Table \ref{table:CoreHFrac} summarizes the opacity multipliers used for the evolution models discussed in this paper, as well as the core hydrogen mass fraction and age for the T$_{\rm eff}$  = 7600 K models.  The opacity multipliers have the form:

\begin{equation}
\kappa = \kappa_{0}(1.0 + M~e^{-[(T - T_{c})/W)]^{2}})
\end{equation}

\noindent
where $T_{c}$ is the center of the opacity bump, and W is the width parameter.  We chose the width so that the bump enhancement would smoothly disappear at the edge of the original bump.  We did not explore the effects of variations of the bump width.  Larger widths could enhance the effects, but are perhaps unrealistic as there are no obvious atomic processes outside the bump regions to motivate opacity enhancements.

In general, for fixed 7600 K effective temperature, the core hydrogen mass fraction becomes larger and ages decrease when opacities are increased in the deepest layers.

\begin{table}[H]
\caption{Opacity bump multiplier parameters, core hydrogen mass fraction and age for 2 M$_{\odot}$ main-sequence models with T$_{\rm eff}$ = 7600 K.}
\centering
\begin{tabular}{lccccc}
\toprule
\textbf{Model}	& \textbf{Bump} &\textbf{Bump}	& \textbf{Bump Width} & \textbf{Core H Fraction} & \textbf{Age (Myr)}\\
                         & \textbf{Coefficient $M$} & \textbf{Center $T_{c}$ (K)}      & \textbf{Parameter $W$ (K)}       &       &       \\
\midrule
OPAL	& ---	& ---	& ---		& 0.2075 & 788.4\\
OPLIB	& ---	& ---	& ---		& 0.2085 & 767.7\\
$\times$2 Hy bump	& 1	& 13,000	& 8,000		& 0.2082 & 767.4 \\
$\times$2 He+ bump	& 1	& 50,000	& 12,000		& 0.1997 & 777.1 \\
$\times$2 Z-bump	& 1	& 200,000 & 30,000	& 0.2097 & 766.2\\
$\times$5 Z-bump	& 4	& 200,000	 & 30,000		& 0.2027 & 773.6 \\
$\times$2 deeper bump & 1		& 350,000		& 30,000	& 0.2103  & 765.8\\
$\times$1.5 deep bump & 0.5		& 2.6$\times$10$^{6}$	& 875,000 & 0.2536 & 720.8\\

\bottomrule
\end{tabular}
  \label{table:CoreHFrac}
\end{table}

\section{Turbulent broadening estimates}
 \label{sect:estimates}

As alluded to throughout this paper, the problem with invoking turbulent broadening to enhance opacities for stellar pulsation studies is that many types of pulsations are driven by the $\kappa$ (opacity valving) mechanism, which requires radiation instead of convection to transport the luminosity in the layer of interest.  However, turbulence/convection must transport at least some of the luminosity in the pulsation driving layer in order for turbulent broadening to affect the opacity.

We carried out some preliminary calculations of the effects of turbulent broadening on Rosseland mean opacity in the region of the Z-bump produced by bound-bound transitions of Fe-group elements in A and B-type main-sequence stars.  For the solar abundance mixture,  turbulent velocity 4$\times$10$^{5}$ cm/s, temperatures 15 to 20 eV (175,000 - 232,000 K), and densities 4.6$\times$10$^{-6}$ to 8.0$\times$10$^{-6}$ g/cm$^{3}$, the Rosseland mean opacity increases by only 0.6\%, far from the factors of nearly two proposed to solve some problems with B-type star pulsations.  However, if the Fe abundance is enhanced at these temperatures because of diffusive element settling and radiative levitation, the opacity increases could be larger.  

We estimate the increase in Doppler broadened line widths taking into account turbulent line broadening for a Fe line in the Z-bump region of the stellar interior.  The thermal width is given by the standard formula:

\begin{equation}
W_{thermal} = \sqrt{2kT/Mc^2}~f_0
\end{equation}

\noindent
where $T$ is temperature, $k$ is Boltzmann's constant, $M$ is the atomic mass of the element, $c$ is the speed of light, and $f_0$ is the frequency of the line center of interest.  The turbulent width is given by the formula:

\begin{equation}
W_{turb} = (v_{turb}/c)~f_0
\end{equation}

\noindent
where $v_{turb}$ is the turbulent velocity and $c$ is the speed of light.  To obtain the Doppler width, these two terms are added in quadrature:

\begin{equation}
W_{Doppler} = \sqrt{W_{thermal}^2 + W_{turb}^2}
\end{equation}

\noindent
Using $k$ = 8.617$\times$10$^{-5}$ eV/K, $T$ = 200,000 K, $v_{turb}$ = 4$\times$10$^{5}$ cm/s, Fe atomic mass 56, mass per nucleon 931.5$\times$10$^{6}$ eV/c$^{2}$, and speed of light 3$\times$10$^{10}$ cm/s,

\begin{equation}
W_{thermal} = 2.571 \times 10^{-5}~f_0
\end{equation}

\begin{equation}
W_{turb} = 1.333 \times 10^{-5}~f_0
\end{equation}

\begin{equation}
W_{Doppler} = 2.896 \times 10^{-5}~f_0
\end{equation}

\noindent
Therefore turbulent broadening would produce a 13\% increase in the Doppler width of a given Fe line. For the A-type stellar models studied here, a turbulent velocity of 4$\times$10$^{5}$ cm/s is already twice as high as the turbulent velocity produced in the Z-bump region with a $\times$5 opacity enhancement.  For a smaller turbulent velocity of 2$\times$10$^{5}$ cm/s, the Doppler width is increased by only 3\%.  These turbulent velocities are calculated by mixing-length theory, with a mixing-length to pressure scale height ratio of 2.0.  A more sophisticated treatment of turbulence and multidimensional models would show a distribution of turbulent velocities, with significantly higher local velocities.

Other elements in the abundance mixture that contribute to opacity in stellar radiative interiors, e.g., C, N, O, and Ne, are more abundant, but are significantly lighter than Fe. Therefore, the enhancement due to turbulence is expected to be less because of the 1/$\sqrt{M}$ dependence in the thermal broadening versus the constant value of $v_{turb}$ for all elements.

This estimate does not directly indicate the corresponding expected increase in Rosseland mean opacity for the stellar abundance mixture, of which Fe is only one element.  Detailed opacity calculations are required, taking into account all of the lines and full abundance mixture under various conditions found in stellar interiors in order to determine whether turbulent broadening can be neglected to a good approximation.  We have yet to estimate the potential effects of edge blending on opacity increases for the regions of ionization of H and He, where the turbulent velocities are higher.

\section{Conclusions}

A-type stars may turn out to be useful testbeds for opacity studies.  Opacity increases could cause convection zones to appear or widen, abundance gradients to be altered, instability strip boundaries to change, unstable mode frequency ranges to change, and could affect operation of pulsation driving mechanisms such as the $\kappa$ effect and mechanisms involving pulsation-convection interactions.  Although angular momentum transport is not investigated in the 1-D non-rotating evolution models presented here, new or wider convection zones and higher convective velocities produced by opacity enhancements could also affect angular momentum transport during stellar evolution and change interior rotation profiles that may have detectable signatures in pulsation frequency spacings.

The tests presented here show that opacity increases proposed to explain discrepancies with pulsations in B-type stars have a more modest effect on pulsations of A-type stars.  With the possible exception of enhancements around the Kurucz bump not explored here, opacity increases do not help to explain the unexpected $g$-mode pulsations discovered in many A-type stars.  However, opacity enhancements in the hydrogen and 1st helium ionization region near the stellar surface may result in driving of additional $p$ modes, confirming that improved treatments of the surface layers may explain the frequency content of A-type stars such as HD 187547 and HR 7284.  More work is warranted to explore the effects of opacities and opacity enhancements on the pulsation instability region boundaries for $\delta$ Sct $p$-mode pulsators.

Convective velocities in pulsation driving layers reach 10$^{5}$-10$^{6}$ cm/s may produce turbulent line and edge broadening, and therefore may need to be taken into account in opacity calculations. However, our estimates show that Rosseland mean opacity increases, taking into account turbulent broadening, are likely to be modest, only around 1\%.

\acknowledgments{We thank P. Walczak for calculating the OPLIB opacity tables with the AGSS09 mixture used in these studies.  We also thank P. Walczak, Stan Owocki, J. Daszy{\'n}ska-Daszkiewicz, and Dietrich Baade for useful discussions.  We also thank the referees for suggestions and questions that greatly improved this paper.  We acknowledge the Department of Energy High Energy Density Physics Impact program for support for this research.}

\authorcontributions{Chris Fryer suggested the idea to investigate turbulent line broadening to enhance opacities and helped to guide this research.  Christopher J. Fontes estimated the effects of turbulent broadening for conditions in stellar interiors.  Joyce A. Guzik decided to investigate the effects on A-type stars, designed the opacity bump studies, performed the stellar evolution and pulsation analyses, and wrote the text and created figures for this article.}

\conflictsofinterest{The authors declare no conflict of interest.  The founding sponsors had no role in the design of the study; in the collection, analyses, or interpretation of data; in the writing of the manuscript, and in the decision to publish the results.} 





\externalbibliography{yes}
\bibliography{Guzik}

\begin{thebibliography}{-------}
\providecommand{\natexlab}[1]{#1}

\bibitem[{Walczak} \em{et~al.}(2015){Walczak}, {Fontes}, {Colgan}, {Kilcrease},
  and {Guzik}]{2015A&A...580L...9W}
{Walczak}, P.; {Fontes}, C.J.; {Colgan}, J.; {Kilcrease}, D.P.; {Guzik}, J.A.
\newblock {Wider pulsation instability regions for {$\beta$} Cephei and SPB
  stars calculated using new Los Alamos opacities}.
\newblock {\em \aap} {\bf 2015}, {\em 580},~L9.
\newblock
  doi:{\changeurlcolor{black}\href{https://doi.org/10.1051/0004-6361/201526824}{\detokenize{10.1051/0004-6361/201526824}}}.

\bibitem[{Walczak} \em{et~al.}(2017){Walczak}, {Daszy{\'n}ska-Daszkiewicz},
  {Pamyatnykh}, {Handler}, {Pigulski}, and {BEST}]{2017arXiv170101258W}
{Walczak}, P.; {Daszy{\'n}ska-Daszkiewicz}, J.; {Pamyatnykh}, A.; {Handler},
  G.; {Pigulski}, A.; {BEST}.
\newblock {Interpretation of the BRITE oscillation spectra of the early B-type
  stars: $\nu$ Eri and $\alpha$ Lup}.
\newblock {\em ArXiv e-prints} {\bf 2017},
  \href{http://xxx.lanl.gov/abs/1701.01258}{{\normalfont
  [arXiv:astro-ph.SR/1701.01258]}}.

\bibitem[{Daszy{\'n}ska-Daszkiewicz}
  \em{et~al.}(2017){Daszy{\'n}ska-Daszkiewicz}, {Pamyatnykh}, {Walczak},
  {Colgan}, {Fontes}, and {Kilcrease}]{2017MNRAS.466.2284D}
{Daszy{\'n}ska-Daszkiewicz}, J.; {Pamyatnykh}, A.A.; {Walczak}, P.; {Colgan},
  J.; {Fontes}, C.J.; {Kilcrease}, D.P.
\newblock {Interpretation of the BRITE oscillation data of the hybrid pulsator
  {$\nu$} Eridani: a call for the modification of stellar opacities}.
\newblock {\em \mnras} {\bf 2017}, {\em 466},~2284--2293,
  \href{http://xxx.lanl.gov/abs/1612.05820}{{\normalfont
  [arXiv:astro-ph.SR/1612.05820]}}.
\newblock
  doi:{\changeurlcolor{black}\href{https://doi.org/10.1093/mnras/stw3315}{\detokenize{10.1093/mnras/stw3315}}}.

\bibitem[{Moravveji}(2016)]{2016MNRAS.455L..67M}
{Moravveji}, E.
\newblock {The impact of enhanced iron opacity on massive star pulsations:
  updated instability strips}.
\newblock {\em \mnras} {\bf 2016}, {\em 455},~L67--L71,
  \href{http://xxx.lanl.gov/abs/1509.08652}{{\normalfont
  [arXiv:astro-ph.SR/1509.08652]}}.
\newblock
  doi:{\changeurlcolor{black}\href{https://doi.org/10.1093/mnrasl/slv142}{\detokenize{10.1093/mnrasl/slv142}}}.

\bibitem[{Cugier}(2014)]{2014A&A...565A..76C}
{Cugier}, H.
\newblock {Seismic models of O and B stars with Kurucz's opacity data}.
\newblock {\em \aap} {\bf 2014}, {\em 565},~A76.
\newblock
  doi:{\changeurlcolor{black}\href{https://doi.org/10.1051/0004-6361/201220507}{\detokenize{10.1051/0004-6361/201220507}}}.

\bibitem[{Zdravkov} and {Pamyatnykh}(2009)]{2009AIPC.1170..388Z}
{Zdravkov}, T.; {Pamyatnykh}, A.A.
\newblock {Modelling hybrid {$\beta$} Cephei/SPB pulsations: {$\gamma$}
  Pegasi}.
\newblock  American Institute of Physics Conference Series; {Guzik}, J.A.;
  {Bradley}, P.A., Eds.,  2009, Vol. 1170, {\em American Institute of Physics
  Conference Series}, pp. 388--390,
  \href{http://xxx.lanl.gov/abs/0909.4643}{{\normalfont
  [arXiv:astro-ph.SR/0909.4643]}}.
\newblock
  doi:{\changeurlcolor{black}\href{https://doi.org/10.1063/1.3246522}{\detokenize{10.1063/1.3246522}}}.

\bibitem[{Colgan} \em{et~al.}(2016){Colgan}, {Kilcrease}, {Magee}, {Sherrill},
  {Abdallah}, {Hakel}, {Fontes}, {Guzik}, and {Mussack}]{2016ApJ...817..116C}
{Colgan}, J.; {Kilcrease}, D.P.; {Magee}, N.H.; {Sherrill}, M.E.; {Abdallah},
  Jr., J.; {Hakel}, P.; {Fontes}, C.J.; {Guzik}, J.A.; {Mussack}, K.A.
\newblock {A New Generation of Los Alamos Opacity Tables}.
\newblock {\em \apj} {\bf 2016}, {\em 817},~116,
  \href{http://xxx.lanl.gov/abs/1601.01005}{{\normalfont
  [arXiv:astro-ph.SR/1601.01005]}}.
\newblock
  doi:{\changeurlcolor{black}\href{https://doi.org/10.3847/0004-637X/817/2/116}{\detokenize{10.3847/0004-637X/817/2/116}}}.

\bibitem[{Iglesias} and {Rogers}(1996)]{1996ApJ...464..943I}
{Iglesias}, C.A.; {Rogers}, F.J.
\newblock {Updated Opal Opacities}.
\newblock {\em \apj} {\bf 1996}, {\em 464},~943.
\newblock
  doi:{\changeurlcolor{black}\href{https://doi.org/10.1086/177381}{\detokenize{10.1086/177381}}}.

\bibitem[{Asplund} \em{et~al.}(2009){Asplund}, {Grevesse}, {Sauval}, and
  {Scott}]{2009ARA&A..47..481A}
{Asplund}, M.; {Grevesse}, N.; {Sauval}, A.J.; {Scott}, P.
\newblock {The Chemical Composition of the Sun}.
\newblock {\em \araa} {\bf 2009}, {\em 47},~481--522,
  \href{http://xxx.lanl.gov/abs/0909.0948}{{\normalfont
  [arXiv:astro-ph.SR/0909.0948]}}.
\newblock
  doi:{\changeurlcolor{black}\href{https://doi.org/10.1146/annurev.astro.46.060407.145222}{\detokenize{10.1146/annurev.astro.46.060407.145222}}}.

\bibitem[{Cox}(1980)]{1980tsp..book.....C}
{Cox}, J.P.
\newblock {\em {Theory of stellar pulsation}};  1980.

\bibitem[{Dziembowski} and {Pamyatnykh}(2008)]{2008MNRAS.385.2061D}
{Dziembowski}, W.A.; {Pamyatnykh}, A.A.
\newblock {The two hybrid B-type pulsators: {$\nu$} Eridani and 12 Lacertae}.
\newblock {\em \mnras} {\bf 2008}, {\em 385},~2061--2068,
  \href{http://xxx.lanl.gov/abs/0801.2451}{{\normalfont [0801.2451]}}.
\newblock
  doi:{\changeurlcolor{black}\href{https://doi.org/10.1111/j.1365-2966.2008.12964.x}{\detokenize{10.1111/j.1365-2966.2008.12964.x}}}.

\bibitem[{Bailey} \em{et~al.}(2015){Bailey}, {Nagayama}, {Loisel}, {Rochau},
  {Blancard}, {Colgan}, {Cosse}, {Faussurier}, {Fontes}, {Gilleron},
  {Golovkin}, {Hansen}, {Iglesias}, {Kilcrease}, {Macfarlane}, {Mancini},
  {Nahar}, {Orban}, {Pain}, {Pradhan}, {Sherrill}, and
  {Wilson}]{2015Natur.517...56B}
{Bailey}, J.E.; {Nagayama}, T.; {Loisel}, G.P.; {Rochau}, G.A.; {Blancard}, C.;
  {Colgan}, J.; {Cosse}, P.; {Faussurier}, G.; {Fontes}, C.J.; {Gilleron}, F.;
  {Golovkin}, I.; {Hansen}, S.B.; {Iglesias}, C.A.; {Kilcrease}, D.P.;
  {Macfarlane}, J.J.; {Mancini}, R.C.; {Nahar}, S.N.; {Orban}, C.; {Pain},
  J.C.; {Pradhan}, A.K.; {Sherrill}, M.; {Wilson}, B.G.
\newblock {A higher-than-predicted measurement of iron opacity at solar
  interior temperatures}.
\newblock {\em \nat} {\bf 2015}, {\em 517},~56--59.
\newblock
  doi:{\changeurlcolor{black}\href{https://doi.org/10.1038/nature14048}{\detokenize{10.1038/nature14048}}}.

\bibitem[{Vinyoles} \em{et~al.}(2017){Vinyoles}, {Serenelli}, {Villante},
  {Basu}, {Bergstr{\"o}m}, {Gonzalez-Garcia}, {Maltoni}, {Pe{\~n}a-Garay}, and
  {Song}]{2017ApJ...835..202V}
{Vinyoles}, N.; {Serenelli}, A.M.; {Villante}, F.L.; {Basu}, S.;
  {Bergstr{\"o}m}, J.; {Gonzalez-Garcia}, M.C.; {Maltoni}, M.;
  {Pe{\~n}a-Garay}, C.; {Song}, N.
\newblock {A New Generation of Standard Solar Models}.
\newblock {\em \apj} {\bf 2017}, {\em 835},~202,
  \href{http://xxx.lanl.gov/abs/1611.09867}{{\normalfont
  [arXiv:astro-ph.SR/1611.09867]}}.
\newblock
  doi:{\changeurlcolor{black}\href{https://doi.org/10.3847/1538-4357/835/2/202}{\detokenize{10.3847/1538-4357/835/2/202}}}.

\bibitem[{Guzik} \em{et~al.}(2016){Guzik}, {Fontes}, {Walczak}, {Wood},
  {Mussack}, and {Farag}]{2016IAUFM..29B.532G}
{Guzik}, J.A.; {Fontes}, C.J.; {Walczak}, P.; {Wood}, S.R.; {Mussack}, K.;
  {Farag}, E.
\newblock {Sound speed and oscillation frequencies for solar models evolved
  with Los Alamos ATOMIC opacities}.
\newblock {\em IAU Focus Meeting} {\bf 2016}, {\em 29},~532--535,
  \href{http://xxx.lanl.gov/abs/1605.04452}{{\normalfont
  [arXiv:astro-ph.SR/1605.04452]}}.
\newblock
  doi:{\changeurlcolor{black}\href{https://doi.org/10.1017/S1743921316006062}{\detokenize{10.1017/S1743921316006062}}}.

\bibitem[{Buldgen} \em{et~al.}(2017{\natexlab{a}}){Buldgen}, {Salmon}, {Noels},
  {Scuflaire}, {Reese}, {Dupret}, {Colgan}, {Fontes}, {Eggenberger}, {Hakel},
  {Kilcrease}, and {Turck-Chi{\`e}ze}]{2017A&A...607A..58B}
{Buldgen}, G.; {Salmon}, S.J.A.J.; {Noels}, A.; {Scuflaire}, R.; {Reese}, D.R.;
  {Dupret}, M.A.; {Colgan}, J.; {Fontes}, C.J.; {Eggenberger}, P.; {Hakel}, P.;
  {Kilcrease}, D.P.; {Turck-Chi{\`e}ze}, S.
\newblock {Seismic inversion of the solar entropy. A case for improving the
  standard solar model}.
\newblock {\em \aap} {\bf 2017}, {\em 607},~A58,
  \href{http://xxx.lanl.gov/abs/1707.05138}{{\normalfont
  [arXiv:astro-ph.SR/1707.05138]}}.
\newblock
  doi:{\changeurlcolor{black}\href{https://doi.org/10.1051/0004-6361/201731354}{\detokenize{10.1051/0004-6361/201731354}}}.

\bibitem[{Buldgen} \em{et~al.}(2017{\natexlab{b}}){Buldgen}, {Salmon},
  {Godart}, {Noels}, {Scuflaire}, {Dupret}, {Reese}, {Colgan}, {Fontes},
  {Eggenberger}, {Hakel}, {Kilcrease}, and {Richard}]{2017MNRAS.472L..70B}
{Buldgen}, G.; {Salmon}, S.J.A.J.; {Godart}, M.; {Noels}, A.; {Scuflaire}, R.;
  {Dupret}, M.A.; {Reese}, D.R.; {Colgan}, J.; {Fontes}, C.J.; {Eggenberger},
  P.; {Hakel}, P.; {Kilcrease}, D.P.; {Richard}, O.
\newblock {Inversions of the Ledoux discriminant: a closer look at the
  tachocline}.
\newblock {\em \mnras} {\bf 2017}, {\em 472},~L70--L74,
  \href{http://xxx.lanl.gov/abs/1709.00287}{{\normalfont
  [arXiv:astro-ph.SR/1709.00287]}}.
\newblock
  doi:{\changeurlcolor{black}\href{https://doi.org/10.1093/mnrasl/slx139}{\detokenize{10.1093/mnrasl/slx139}}}.

\bibitem[{Balona} \em{et~al.}(2015){Balona}, {Daszy{\'n}ska-Daszkiewicz}, and
  {Pamyatnykh}]{2015MNRAS.452.3073B}
{Balona}, L.A.; {Daszy{\'n}ska-Daszkiewicz}, J.; {Pamyatnykh}, A.A.
\newblock {Pulsation frequency distribution in {$\delta$} Scuti stars}.
\newblock {\em \mnras} {\bf 2015}, {\em 452},~3073--3084,
  \href{http://xxx.lanl.gov/abs/1505.07216}{{\normalfont
  [arXiv:astro-ph.SR/1505.07216]}}.
\newblock
  doi:{\changeurlcolor{black}\href{https://doi.org/10.1093/mnras/stv1513}{\detokenize{10.1093/mnras/stv1513}}}.

\bibitem[{Seaton}(2005)]{2005MNRAS.362L...1S}
{Seaton}, M.J.
\newblock {Opacity Project data on CD for mean opacities and radiative
  accelerations}.
\newblock {\em \mnras} {\bf 2005}, {\em 362},~L1--L3,
  \href{http://xxx.lanl.gov/abs/astro-ph/0411010}{{\normalfont
  [astro-ph/0411010]}}.
\newblock
  doi:{\changeurlcolor{black}\href{https://doi.org/10.1111/j.1365-2966.2005.00019.x}{\detokenize{10.1111/j.1365-2966.2005.00019.x}}}.

\bibitem[{Castelli} and {Kurucz}(2003)]{2003IAUS..210P.A20C}
{Castelli}, F.; {Kurucz}, R.L.
\newblock {New Grids of ATLAS9 Model Atmospheres}.
\newblock  Modelling of Stellar Atmospheres; {Piskunov}, N.; {Weiss}, W.W.;
  {Gray}, D.F., Eds.,  2003, Vol. 210, {\em IAU Symposium}, p. A20.

\bibitem[{Daszy{\'n}ska-Daszkiewicz}
  \em{et~al.}(2017){Daszy{\'n}ska-Daszkiewicz}, {Walczak}, and
  {Pamyatnykh}]{2017EPJWC.16003013D}
{Daszy{\'n}ska-Daszkiewicz}, J.; {Walczak}, P.; {Pamyatnykh}, A.
\newblock {On possible explanations of pulsations in Maia stars}.
\newblock  European Physical Journal Web of Conferences,  2017, Vol. 160, {\em
  European Physical Journal Web of Conferences}, p. 03013,
  \href{http://xxx.lanl.gov/abs/1701.00937}{{\normalfont
  [arXiv:astro-ph.SR/1701.00937]}}.
\newblock
  doi:{\changeurlcolor{black}\href{https://doi.org/10.1051/epjconf/201716003013}{\detokenize{10.1051/epjconf/201716003013}}}.

\bibitem[{B{\"o}hm-Vitense}(1958)]{1958ZA.....46..108B}
{B{\"o}hm-Vitense}, E.
\newblock {{\"U}ber die Wasserstoffkonvektionszone in Sternen verschiedener
  Effektivtemperaturen und Leuchtkr{\"a}fte. Mit 5 Textabbildungen}.
\newblock {\em \zap} {\bf 1958}, {\em 46},~108.

\bibitem[{Krief} \em{et~al.}(2016){Krief}, {Feigel}, and
  {Gazit}]{2016ApJ...824...98K}
{Krief}, M.; {Feigel}, A.; {Gazit}, D.
\newblock {Line Broadening and the Solar Opacity Problem}.
\newblock {\em \apj} {\bf 2016}, {\em 824},~98,
  \href{http://xxx.lanl.gov/abs/1603.01153}{{\normalfont
  [arXiv:astro-ph.SR/1603.01153]}}.
\newblock
  doi:{\changeurlcolor{black}\href{https://doi.org/10.3847/0004-637X/824/2/98}{\detokenize{10.3847/0004-637X/824/2/98}}}.

\bibitem[{Turcotte} \em{et~al.}(2000){Turcotte}, {Richer}, {Michaud}, and
  {Christensen-Dalsgaard}]{2000A&A...360..603T}
{Turcotte}, S.; {Richer}, J.; {Michaud}, G.; {Christensen-Dalsgaard}, J.
\newblock {The effect of diffusion on pulsations of stars on the upper main
  sequence --- {$\delta$} Scuti and metallic A stars}.
\newblock {\em \aap} {\bf 2000}, {\em 360},~603--616,
  \href{http://xxx.lanl.gov/abs/astro-ph/0006272}{{\normalfont
  [astro-ph/0006272]}}.

\bibitem[{Th{\'e}ado} \em{et~al.}(2012){Th{\'e}ado}, {Alecian}, {LeBlanc}, and
  {Vauclair}]{2012A&A...546A.100T}
{Th{\'e}ado}, S.; {Alecian}, G.; {LeBlanc}, F.; {Vauclair}, S.
\newblock {The new Toulouse-Geneva stellar evolution code including radiative
  accelerations of heavy elements}.
\newblock {\em \aap} {\bf 2012}, {\em 546},~A100,
  \href{http://xxx.lanl.gov/abs/1210.4360}{{\normalfont
  [arXiv:astro-ph.SR/1210.4360]}}.
\newblock
  doi:{\changeurlcolor{black}\href{https://doi.org/10.1051/0004-6361/201219610}{\detokenize{10.1051/0004-6361/201219610}}}.

\bibitem[{Grigahc{\`e}ne} \em{et~al.}(2010){Grigahc{\`e}ne}, {Uytterhoeven},
  {Antoci}, {Balona}, {Catanzaro}, {Daszy{\'n}ska-Daszkiewicz}, {Guzik},
  {Handler}, {Houdek}, {Kurtz}, {Marconi}, {Monteiro}, {Moya}, {Ripepi},
  {Su{\'a}rez}, {Borucki}, {Brown}, {Christensen-Dalsgaard}, {Gilliland},
  {Jenkins}, {Kjeldsen}, {Koch}, {Bernabei}, {Bradley}, {Breger}, {Di
  Criscienzo}, {Dupret}, {Garc{\'{\i}}a}, {Garc{\'{\i}}a Hern{\'a}ndez},
  {Jackiewicz}, {Kaiser}, {Lehmann}, {Mart{\'{\i}}n-Ruiz}, {Mathias},
  {Molenda-{\.Z}akowicz}, {Nemec}, {Nuspl}, {Papar{\'o}}, {Roth}, {Szab{\'o}},
  {Suran}, and {Ventura}]{2010AN....331..989G}
{Grigahc{\`e}ne}, A.; {Uytterhoeven}, K.; {Antoci}, V.; {Balona}, L.;
  {Catanzaro}, G.; {Daszy{\'n}ska-Daszkiewicz}, J.; {Guzik}, J.A.; {Handler},
  G.; {Houdek}, G.; {Kurtz}, D.W.; {Marconi}, M.; {Monteiro}, M.J.P.F.G.;
  {Moya}, A.; {Ripepi}, V.; {Su{\'a}rez}, J.C.; {Borucki}, W.J.; {Brown}, T.M.;
  {Christensen-Dalsgaard}, J.; {Gilliland}, R.L.; {Jenkins}, J.M.; {Kjeldsen},
  H.; {Koch}, D.; {Bernabei}, S.; {Bradley}, P.; {Breger}, M.; {Di Criscienzo},
  M.; {Dupret}, M.A.; {Garc{\'{\i}}a}, R.A.; {Garc{\'{\i}}a Hern{\'a}ndez}, A.;
  {Jackiewicz}, J.; {Kaiser}, A.; {Lehmann}, H.; {Mart{\'{\i}}n-Ruiz}, S.;
  {Mathias}, P.; {Molenda-{\.Z}akowicz}, J.; {Nemec}, J.M.; {Nuspl}, J.;
  {Papar{\'o}}, M.; {Roth}, M.; {Szab{\'o}}, R.; {Suran}, M.D.; {Ventura}, R.
\newblock {Kepler observations: Light shed on the hybrid {$\gamma$} Doradus -
  {$\delta$} Scuti pulsation phenomenon}.
\newblock {\em Astronomische Nachrichten} {\bf 2010}, {\em 331},~989.
\newblock
  doi:{\changeurlcolor{black}\href{https://doi.org/10.1002/asna.201011443}{\detokenize{10.1002/asna.201011443}}}.

\bibitem[{Uytterhoeven} \em{et~al.}(2011){Uytterhoeven}, {Moya},
  {Grigahc{\`e}ne}, {Guzik}, {Guti{\'e}rrez-Soto}, {Smalley}, {Handler},
  {Balona}, {Niemczura}, {Fox Machado}, {Benatti}, {Chapellier}, {Tkachenko},
  {Szab{\'o}}, {Su{\'a}rez}, {Ripepi}, {Pascual}, {Mathias},
  {Mart{\'{\i}}n-Ru{\'{\i}}z}, {Lehmann}, {Jackiewicz}, {Hekker},
  {Gruberbauer}, {Garc{\'{\i}}a}, {Dumusque}, {D{\'{\i}}az-Fraile}, {Bradley},
  {Antoci}, {Roth}, {Leroy}, {Murphy}, {De Cat}, {Cuypers}, {Kjeldsen},
  {Christensen-Dalsgaard}, {Breger}, {Pigulski}, {Kiss}, {Still}, {Thompson},
  and {van Cleve}]{2011A&A...534A.125U}
{Uytterhoeven}, K.; {Moya}, A.; {Grigahc{\`e}ne}, A.; {Guzik}, J.A.;
  {Guti{\'e}rrez-Soto}, J.; {Smalley}, B.; {Handler}, G.; {Balona}, L.A.;
  {Niemczura}, E.; {Fox Machado}, L.; {Benatti}, S.; {Chapellier}, E.;
  {Tkachenko}, A.; {Szab{\'o}}, R.; {Su{\'a}rez}, J.C.; {Ripepi}, V.;
  {Pascual}, J.; {Mathias}, P.; {Mart{\'{\i}}n-Ru{\'{\i}}z}, S.; {Lehmann}, H.;
  {Jackiewicz}, J.; {Hekker}, S.; {Gruberbauer}, M.; {Garc{\'{\i}}a}, R.A.;
  {Dumusque}, X.; {D{\'{\i}}az-Fraile}, D.; {Bradley}, P.; {Antoci}, V.;
  {Roth}, M.; {Leroy}, B.; {Murphy}, S.J.; {De Cat}, P.; {Cuypers}, J.;
  {Kjeldsen}, H.; {Christensen-Dalsgaard}, J.; {Breger}, M.; {Pigulski}, A.;
  {Kiss}, L.L.; {Still}, M.; {Thompson}, S.E.; {van Cleve}, J.
\newblock {The Kepler characterization of the variability among A- and F-type
  stars. I. General overview}.
\newblock {\em \aap} {\bf 2011}, {\em 534},~A125,
  \href{http://xxx.lanl.gov/abs/1107.0335}{{\normalfont
  [arXiv:astro-ph.SR/1107.0335]}}.
\newblock
  doi:{\changeurlcolor{black}\href{https://doi.org/10.1051/0004-6361/201117368}{\detokenize{10.1051/0004-6361/201117368}}}.

\bibitem[{Balona} and {Dziembowski}(2011)]{2011MNRAS.417..591B}
{Balona}, L.A.; {Dziembowski}, W.A.
\newblock {Kepler observations of {$\delta$} Scuti stars}.
\newblock {\em \mnras} {\bf 2011}, {\em 417},~591--601.
\newblock
  doi:{\changeurlcolor{black}\href{https://doi.org/10.1111/j.1365-2966.2011.19301.x}{\detokenize{10.1111/j.1365-2966.2011.19301.x}}}.

\bibitem[{Guzik} \em{et~al.}(2000){Guzik}, {Kaye}, {Bradley}, {Cox}, and
  {Neuforge}]{2000ApJ...542L..57G}
{Guzik}, J.A.; {Kaye}, A.B.; {Bradley}, P.A.; {Cox}, A.N.; {Neuforge}, C.
\newblock {Driving the Gravity-Mode Pulsations in {$\gamma$} Doradus
  Variables}.
\newblock {\em \apjl} {\bf 2000}, {\em 542},~L57--L60.
\newblock
  doi:{\changeurlcolor{black}\href{https://doi.org/10.1086/312908}{\detokenize{10.1086/312908}}}.

\bibitem[{Deal} \em{et~al.}(2016){Deal}, {Richard}, and
  {Vauclair}]{2016A&A...589A.140D}
{Deal}, M.; {Richard}, O.; {Vauclair}, S.
\newblock {Hydrodynamical instabilities induced by atomic diffusion in A stars
  and their consequences}.
\newblock {\em \aap} {\bf 2016}, {\em 589},~A140,
  \href{http://xxx.lanl.gov/abs/1604.01241}{{\normalfont
  [arXiv:astro-ph.SR/1604.01241]}}.
\newblock
  doi:{\changeurlcolor{black}\href{https://doi.org/10.1051/0004-6361/201628180}{\detokenize{10.1051/0004-6361/201628180}}}.

\bibitem[{Deal} \em{et~al.}(2017){Deal}, {Richard}, and
  {Vauclair}]{2017sf2a.conf...31D}
{Deal}, M.; {Richard}, O.; {Vauclair}, S.
\newblock {Hydrodynamical instabilities induced by atomic diffusion in F and A
  stars : Impact on the opacity profile and asteroseimic age determination}.
\newblock  SF2A-2017: Proceedings of the Annual meeting of the French Society
  of Astronomy and Astrophysics; {Reyl{\'e}}, C.; {Di Matteo}, P.; {Herpin},
  F.; {Lagadec}, E.; {Lan{\c c}on}, A.; {Meliani}, Z.; {Royer}, F., Eds.,
  2017, pp. 31--34.

\bibitem[{Antoci} \em{et~al.}(2014){Antoci}, {Cunha}, {Houdek}, {Kjeldsen},
  {Trampedach}, {Handler}, {L{\"u}ftinger}, {Arentoft}, and
  {Murphy}]{2014ApJ...796..118A}
{Antoci}, V.; {Cunha}, M.; {Houdek}, G.; {Kjeldsen}, H.; {Trampedach}, R.;
  {Handler}, G.; {L{\"u}ftinger}, T.; {Arentoft}, T.; {Murphy}, S.
\newblock {The Role of Turbulent Pressure as a Coherent Pulsational Driving
  Mechanism: The Case of the {$\delta$} Scuti Star HD 187547}.
\newblock {\em \apj} {\bf 2014}, {\em 796},~118,
  \href{http://xxx.lanl.gov/abs/1411.0931}{{\normalfont
  [arXiv:astro-ph.SR/1411.0931]}}.
\newblock
  doi:{\changeurlcolor{black}\href{https://doi.org/10.1088/0004-637X/796/2/118}{\detokenize{10.1088/0004-637X/796/2/118}}}.

\bibitem[{Samadi} \em{et~al.}(2002){Samadi}, {Goupil}, and
  {Houdek}]{2002A&A...395..563S}
{Samadi}, R.; {Goupil}, M.J.; {Houdek}, G.
\newblock {Solar-like oscillations in delta Scuti stars}.
\newblock {\em \aap} {\bf 2002}, {\em 395},~563--571,
  \href{http://xxx.lanl.gov/abs/astro-ph/0208573}{{\normalfont
  [astro-ph/0208573]}}.
\newblock
  doi:{\changeurlcolor{black}\href{https://doi.org/10.1051/0004-6361:20021322}{\detokenize{10.1051/0004-6361:20021322}}}.

\end{thebibliography}

\end{document}